\documentclass[11pt]{article}
\usepackage{amsmath,amssymb,color,graphics,epsfig,bm}
\usepackage{graphicx}
\usepackage{subfigure}

\usepackage{amsfonts}

\newcommand{\be}{\begin{equation}}
\newcommand{\ee}{\end{equation}}
\newcommand{\bea}{\setlength\arraycolsep{2pt} \begin{eqnarray}}
\newcommand{\eea}{\end{eqnarray}}
\newcommand{\nn}{\nonumber}

\newcommand{\oor}{\text{\uppercase\expandafter{\romannumeral1}}}
\newcommand{\tr}{\text{\uppercase\expandafter{\romannumeral2}}}
\newcommand{\ttr}{\text{\uppercase\expandafter{\romannumeral3}}}
\newcommand{\fr}{\text{\uppercase\expandafter{\romannumeral4}}}

\def\ft#1#2{{\textstyle{\frac{\scriptstyle #1}{\scriptstyle #2} } }}
\def\fft#1#2{{\frac{#1}{#2}}}

\def\0{{\sst{(0)}}}
\def\1{{\sst{(1)}}}
\def\2{{\sst{(2)}}}
\def\3{{\sst{(3)}}}
\def\4{{\sst{(4)}}}
\def\5{{\sst{(5)}}}
\def\6{{\sst{(6)}}}
\def\7{{\sst{(7)}}}
\def\8{{\sst{(8)}}}
\def\sst#1{{\scriptscriptstyle #1}}

\begin{document}

\begin{flushright}
\end{flushright}

\vspace{25pt}
\begin{center}
{\large {\bf Holographic complexity and thermodynamics of AdS black holes}}

\vspace{10pt}
 Zhong-Ying Fan$^1$, Minyong Guo$^{2,3}$\\

\vspace{10pt}
$^1${ Center for Astrophysics, School of Physics and Electronic Engineering, \\
 Guangzhou University, Guangzhou 510006, China }\\
 $^2${ Department of Physics, Beijing Normal University, \\
 Beijing 100875,  P. R. China}\\
$^3${ Perimeter Institute for Theoretical Physics \\Waterloo, Ontario N2L 2Y5, Canada\\}
\smallskip

\vspace{40pt}

\underline{ABSTRACT}
\end{center}
In this paper, we relate the complexity for a holographic state to a simple gravitational object of which the growth rate at late times is equal to temperature times black hole entropy. We show that if this is correct, the thermodynamics of AdS black holes implies that for generic holographic states dual to static AdS black holes, the complexity growth rate at late times will saturate the Lloyd bound at high temperature limit. In particular, for AdS planar black holes, the result holds at lower temperatures as well. We conjecture that the complexity growth is bounded above as $d\mathcal{C}/dt\leq \alpha T S/\pi\hbar$ or $d\mathcal{C}/dt\leq \alpha \big(T_+ S_+-T_-S_-\big)/\pi\hbar$ for black holes with an inner horizon, where $\alpha$ is an overall coefficient for our new proposal. The conjecture passes a number of nontrivial tests for black holes in Einstein's gravity. However, we also find that the bound may be violated in the presence of stringy corrections.

\vfill {\footnotesize  Email: fanzhy@gzhu.edu.cn\,,\quad guominyong@gmail.com\,.}

\thispagestyle{empty}

\pagebreak

\tableofcontents
\addtocontents{toc}{\protect\setcounter{tocdepth}{2}}




\section{Introduction}

The speed of computation is a central topic in the field of quantum computations. The importation notion that characterizes the computational speed of a quantum computer is called complexity. Complexity is defined by the minimal number of elementary operations (or logic gates) needed to build a target state of interest from a given reference state. Intuitively, one expects that any way to produce the target state through a given quantum circuit has already led to an upper bound on complexity. However, what the bound is remains a great mystery.

Inspired by the Margolus-Levitin theorem \cite{Margolus:1997ih}, it was conjectured by Lloyd that \cite{Lloyd}
\be \fft{d\mathcal{C}}{dt}\leq \fft{2\langle E \rangle}{\pi\hbar} \,,\ee
where $\langle E \rangle$ is the energy of the target state and the ground energy is taken to be zero. The bound is fascinating since it is simple and powerful. However, whether the computational speed is really limited by the energy considerations alone for a system was recently questioned in \cite{Jordan:2017vqh}. It was argued there that the energy alone is not sufficient to derive an upper bound on the computational speed. Instead, additional physical conditions, for example information density and information transmission speed should be considered appropriately.

On the other hand, recent developments in AdS/CFT correspondence give rise to several distinct proposals for complexity for a holographic state \cite{Stanford:2014jda,Brown:2015bva,Brown:2015lvg,Couch:2016exn,Fan:2018wnv}. However, though these proposals show interesting properties of complexity and have attracted a lot of attentions in literature \cite{Fan:2018xwf,Moosa:2017yiz,HosseiniMansoori:2017tsm,Mahapatra:2018gig,Chapman:2016hwi,Carmi:2016wjl,Kim:2017lrw,Yang:2016awy,
Moosa:2017yvt,Swingle:2017zcd,Alishahiha:2018tep,An:2018xhv,Jiang:2018pfk,Kim:2017qrq,Yang:2019gce,Guo:2019vni,Cai:2016xho,Lehner:2016vdi,Huang:2016fks,Cano:2018aqi,Jiang:2018sqj,Jiang:2019fpz,Feng:2018sqm,Alishahiha:2017hwg,Carmi:2017jqz}, none of them obeys the Lloyd bound for general holographic theories. It becomes an open question that whether there exists a universal upper bound on the growth rate of complexity.

In this paper, we would like to some extent to address a related issue: for what theories (states) or under what conditions, the Lloyd bound is valid. For this purpose, we study the complexity for quantum field theories at strong coupling via gauge/gravity duality. Based on an argument in \cite{Susskind:2014rva}, we propose that the complexity for a holographic state is dual to a simple gravitational object defined on Wheeler-DeWitt (WDW) patch, of which the growth rate at late times is equal to temperature times black hole entropy. We show that if this is correct, the thermodynamics of AdS black holes guarantees that that the Lloyd bound naturally emerges at late times for generally static AdS black holes with spherical/hyperbolic/toric symmetries at high temperature regimes. In particular, for planar black holes, the result holds at lower temperatures as well. Inspired by universality of the result, we conjecture that the complexity growth is bounded above as $d\mathcal{C}/dt\leq \alpha T S/\pi\hbar$ or $d\mathcal{C}/dt\leq \alpha \big(T_+ S_+-T_-S_-\big)/\pi\hbar$ for black holes with an inner horizon, where $\alpha$ is an overall coefficient for our new proposal. To test the conjecture, we study the time dependence of complexity for holographic states without perturbed shock waves. As far as we can check, the bound is valid to black holes in Einstein's gravity if matter fields satisfy strong and weak energy conditions. However, it may be violated in the presence of stringy corrections.

The remaining of the paper is organized as follows. In section 2, we study the complexity growth rate at late times from thermodynamics of AdS black holes. In section 3, we construct a new gravitational object of which the growth rate at late times is equal to temperature times black hole entropy. We then introduce our new proposal for holographic complexity. In section 4, we study the time dependence of complexity for a variety of black holes. In section 5, we study the growth rate of an alternative object: the joint action of WDW patch which gives the same result at late times. We conclude in section 6.

\section{Universal results about complexity growth at late times}

Based on a quantum circuit model, it was argued in \cite{Susskind:2014rva} that apart from a early transit regime, the rate of change of complexity is given by entropy $S$ times temperature $T$ (in geometric units): $d\mathcal{C}/dt\sim T S/\pi\hbar$. The entropy characterizes the width of the circuit whilst the temperature (or precisely speaking its inverse $\hbar/k_B T$) describes the local rate at which a certain qubit interacts. Indeed, the result is natural for high temperature systems, such as black holes. It has also been verified for various distinct proposals for holographic complexity \cite{Stanford:2014jda,Brown:2015bva,Brown:2015lvg,Couch:2016exn,Fan:2018wnv}. However, a major shortcoming of these proposals is the overall coefficient is not universal for all of them.

Motivated by this, we would like to search a gravitational object of which the rate of change at late times is equal to $TS$, even at low temperature regimes, so that we can choose the coefficient to be universal (this is said in the sense that the states under considerations live in a same dimensional spacetime). In other words, the complexity growth at late times will be given by
\be\label{lategrowth} \fft{d\mathcal{C}}{dt}=\fft{\alpha T S}{\pi\hbar} \,,\ee
where $\alpha$ is an overall coefficient introduced by our new proposal (see Eq.(\ref{proposal1})). For later purpose, we take it to be $\alpha=\ft{2(D-2)}{D-1}$, where $D$ denotes spacetime dimension.

The construction of new proposals for complexity obeying Eq.(\ref{lategrowth}) will be studied in details in sec.\ref{newproposal} and sec.\ref{altenateproposal}.
In this section, we shall first examine to what extent the growth rate of complexity at late times is universal for any such proposal. For this purpose, we recall that for generally static AdS planar black holes
\be ds^2=-h(r)dt^2+dr^2/f(r)+r^2 dx^idx^i \,,\ee
there exists an extra scaling symmetry
\be\label{scalingsymmetry} r\rightarrow \lambda r\,,\quad(t\,,x^i)\rightarrow \lambda^{-1} (t\,,x^i)\,,\quad(h\,,f)\rightarrow \lambda^2 (h\,,f)\,.\ee
It leads to new scaling behaviors for thermodynamic quantities. For example, for a neutral black hole with a first law $dM=T dS$, one has
\be\label{scaling} M\rightarrow \lambda^{D-1}M\,,\quad T\rightarrow \lambda T\,,\quad S\rightarrow \lambda^{D-2}S\,.\ee
Then a standard scaling dimensional argument will lead to a generalised Smarr formula \cite{Liu:2015tqa}
\be\label{smarr0} M=\ft{D-2}{D-1}\, TS\,.\ee
In the presence of matter fields, for example, a scalar field, the situation is a little more evolved because the first law will be modified by some nonconserved (scalar) charges \cite{Lu:2014maa}. However, a careful examination in \cite{Liu:2015tqa} shows that the above generalised Smarr relation continues hold, with the mass $M$ defined by the usually massless graviton mode (the fall-off mode $1/r^{D-3}$ at asymptotic infinity). For general matter fields (not gauge fields), we expect that there will always exist an energy function (we still denote it by $M$ without confusion), which obeys the relation (\ref{smarr0}) since the existence of such a Smarr-like relation is simply associated with the global scaling symmetry (\ref{scalingsymmetry}) of AdS planar black holes. While $M$ may not be the mass of the solutions\footnote{By ``mass'', we mean the energy defined by the usually massless graviton mode. In general, $M$ is related to the mass via a Legendre transformation}, this does not introduce any shortcomings for our purpose since it is still an energy function of the state under considerations. Therefore, with our choice for the coefficient $\alpha$, the rate of change of complexity at late times manifestly saturates the Lloyd bound $ d\mathcal{C}/dt=2 M/\pi\hbar$. Remarkably, this is valid to generally neutral AdS planar black holes of any size, in any number of dimensions and with or without nonconserved charges!

For charged black holes, the electrostatic potential $\mu$ (electric charge $Q$) scales in the same way of the temperature $T$ (entropy $S$). Hence, one has
\be\label{smarr1} M=\ft{D-2}{D-1} \big( TS+\mu Q \big)\,.\ee
It follows that $d\mathcal{C}/dt=2 (M-\tilde\mu Q)/\pi\hbar$, where $\tilde \mu=\ft{D-2}{D-1}\,\mu$. Interestingly, the result is slightly bigger than the original expectation $2(M-\mu Q)/\pi\hbar$ in \cite{Brown:2015lvg}. However, it does not mean that it conflicts with the latter. In fact, for a given system with a fixed charge (or chemical potential), the original scaling $\mu\rightarrow \lambda \mu\,,Q\rightarrow \lambda^{D-2}Q$ is clearly breaking. Indeed, in high temperature limit, the Smarr formula to leading order gives (the derivation is given in Appendix \ref{smarrhightem})
\be\label{smarr2} M-\mu Q\simeq \ft{D-2}{D-1}\,T S \,.\ee
This will reproduce the expected result in \cite{Brown:2015lvg} for complexity growth rate of charged systems. The enhancement of upper bound on computation rate at lower temperatures may be attributed to weaker thermal barrier and stronger (non-thermal) interactions for charged carriers: they get more useful energy and process quantum information faster than they do at high temperatures. The effective result is it looks like that the chemical potential and electric charge increase (decrease) as the temperature increases (decreases) in the manner of $\mu\propto T\,,Q\propto T^{D-2}\,,$ at lower temperature regimes. It is of great interests to explore whether and how this phenomenon emerges in the context of quantum field theories.

 For charged black holes with an inner horizon, the generalised Smarr relation (\ref{smarr1}) holds on both of the horizons. The rate of change of complexity at late times is given by
\bea\label{cpinner}
\fft{d\mathcal{C}}{dt}&=&\fft{2}{\pi\hbar}\Big[\big(M- \tilde{\mu}Q\big)_+-\big(M-\tilde{\mu} Q \big)_-\Big] \nn\\
&=& \fft{2}{\pi\hbar}\Big[\big(M- \mu Q\big)_+-\big(M-\mu Q \big)_-\Big]_{T\rightarrow \infty}\,.
\eea
Up to an overall coefficient, the upper bound of this type was first conjectured in \cite{Cai:2016xho}. However, the subtracted contribution on the inner horizon lacks a field theory description. Nevertheless, in the context of AdS/CFT correspondence, the computation rate for black holes with an inner horizon indeed should be reduced since a part of microscopic degrees of freedoms inside the black hole does not take part in mixing quantum information. The inner horizon looks like a cutoff (or a gap) from the point of view of processing information.

Next, we consider stringy corrections to complexification rate. An interesting example is the Gauss-Bonnet black hole (\ref{GBBH}) with a positive coupling constant. In this case, the generalised Smarr relation (\ref{smarr0}) continues hold  \cite{Liu:2015tqa}. However, the computation rate at late times is in fact reduced, compared with that for black holes in Einstein's gravity with a same massless graviton mode. This is because stringy corrections reduces the black hole mass as
\be M=\sqrt{1-4\lambda}\,M_{E}<M_E \,.\ee
This is consistent with the expectations in \cite{Brown:2015lvg} that stringy corrections should reduce the computation rate of black holes.

For spherical/hyperbolic black holes, the scaling symmetry (\ref{scalingsymmetry}) is explicitly breaking owing to nontrivial topology of the black holes so that the generalised Smarr formula (\ref{smarr1}) does not hold any longer. Hence, in this case there are not universal results about complexity growth at low temperatures. However, in high temperature limit, the result (\ref{smarr2}) holds as the planar case. The reason is simple: for very large black holes, the topology becomes less important. In other words, the solutions are approximately planar. In fact, it is known that a static planar black hole can be obtained from its spherical counterpart by using the scaling (\ref{scalingsymmetry}) and sending $\lambda\rightarrow \infty$. This is in the same spirit of taking high temperature limit.

In conclude, if there exists a proposal for complexity satisfying (\ref{lategrowth}), then the growth rate at late times will saturate the Lloyd bound and its proper generalisations for charged systems at high temperature limit. This is universal to holographic states described by generally static AdS black holes with spherical/hyperbolic/toric isometries.

In the subsequent section, we will show how to construct a gravitational object of which the rate of change at late times is equal to (or bounded above by) $T S$ and then introduce a new proposal for holographic complexity.

\section{A new proposal for holographic complexity}\label{newproposal}
 The product $TS$ looks familiar for gravity theorists. It was first shown by Wald \cite{wald1,wald2} that for any gravitational theory with diffeomorphism invariance, a geometric definition for black holes entropy can be introduced as follows
\be TS=\int_{r=r_h} \bm{Q}_{grav} \,,\ee
where boldface letters denote form fields and $\bm{Q}_{grav}$ stands for the Noether charge associated to a time-like Killing vector for gravity sector (relative to matter sector). However, what we need now is not a single object valid to stationary black holes but a diffeomorphism invariant quantity. A careful examination of this will lead to our new proposal for holographic complexity.

Let us first review how the Noether charge is introduced in the Wald-Iyer formalism. The variation of the Lagrangian gives
\be \delta \mathbf{L}=\mathbf{E}\,\delta \psi+d\mathbf{\Theta} \,,\ee
where $\psi$ collectively denotes all the dynamical fields and $\mathbf{\Theta}=\mathbf{\Theta}(\psi\,,\delta \psi)$ is the presymplectic current form. For a diffeomorphism invariant theory, any vector field $\bm\xi$ defined on the curved manifold constitutes an infinitesimally local symmetry for the solutions. Hence, to each $\bm\xi$, one can define a Noether current as
\be\label{noethercurrent1} \mathbf J\equiv \mathbf\Theta(\psi\,,\mathcal{L}_\xi\psi)-\bm\xi\cdot \mathbf L \,.\ee
The Noether charge $\mathbf Q$ is defined by $\mathbf J=d\mathbf Q$ because $\mathbf J$ is closed for any on-shell solutions \cite{wald1,wald2}. One finds
\be d\mathbf{Q}=\mathbf\Theta(\psi\,,\mathcal{L}_\xi\psi)-\bm\xi\cdot \mathbf L \,.\ee
The dual form of this equation gives
\be *d*\bm{\mathcal{Q}}=**\bm{j}-*\big(\bm\xi\cdot *\mathcal{L} \big) \,,\label{central1}\ee
where $\bm{\mathcal{Q}}$ is the 2-form Noether charge, $\bm j$ is the one-form presympletic current and $\mathcal{L}$ is the Lagrangian density. More explicitly, one has
\be \nabla_\nu \mathcal{Q}^{\mu\nu}=j^\mu-\xi^\mu\mathcal{L} \,.\label{central2}\ee
It should be emphasized that this is valid for any vector field $\xi$. In particular, when $\xi$ is a Killing vector field, one has $\delta \psi=L_\xi \psi=0$, leading to $j^\mu(\delta\psi)=0$ since the current $j^\mu$ depends linearly on $\delta \psi$.

However, for our current purpose, the equation (\ref{central2}) is oversimplified. What we need is just a similar equation valid to the Noether charge for gravity sector. It was established in \cite{Fan:2018qnt} that for general higher derivative gravities $\mathcal{L}_{grav}=\mathcal{L}_{grav}(g_{\mu\nu};R_{\mu\nu\rho\sigma})$, one has
\bea
\nabla_\nu \mathcal{Q}^{\mu\nu}_{grav}=j^\mu_{grav}-\xi^\mu \mathcal{L}_{grav}-2\xi_\nu T^{\mu\nu}\,,
\eea
where $T_{\mu\nu}$ stands for the energy-momentum tensor of matter fields. It follows that for a time-like Killing vector $\xi=\partial/\partial t$, one finds for static black holes
\be\label{bulk} \partial_r\Big(\sqrt{-g}\mathcal{Q}_{grav}^{rt} \Big)=\sqrt{-g}\,\big( \mathcal{L}_{grav}-2\rho\big) \,,\ee
where $\rho=-T^t_{\,\,t}$ is the local energy density of matter fields. This leads to
\bea\label{waldidentity} T S&= &\int_{r=r_h}\bm{Q}_{grav}\nn\\
&=&\int d\Omega_{(D-2)}\Big(-\sqrt{-g}\,\mathcal{Q}^{rt}_{grav} \Big)_{r=r_h} \nn\\
&=&-\int d\Omega_{(D-2)}\int_{BH} d r\,\sqrt{-g}\,\big( \mathcal{L}_{grav}-2\rho\big)\nn\\
&&+\int d\Omega_{(D-2)}\Big(-\sqrt{-g}\,\mathcal{Q}^{rt}_{grav} \Big)_{r=\epsilon}\,,\eea
where in the third line, we have adopted the relation (\ref{bulk}). It is worth emphasizing that the contribution near singularity is rather subtle. In the semi-classical limit, the second term on the r.h.s of (\ref{waldidentity}) might be nonzero\footnote{In fact, with stringy corrections, the Noether charge may diverge for certain cases, for example the Gauss-Bonnet black holes with $\lambda<0$, which have an alternate singularity at some finite radii. The same situation happens for the ``Complexity=Action" proposal \cite{Brown:2015bva,Brown:2015lvg}. Of course, it cannot be treated credibly without a well developed quantum gravity theory.}. However, as will be shown later, the Noether charge term evaluated at the singularity is always nonnegative (at least for Einstein gravity) in the semi-classical limit, namely $\Big(-\sqrt{-g}\,\mathcal{Q}^{rt}_{grav} \Big)_{r=\epsilon}\geq 0$. Thus, one finds from (\ref{waldidentity})
\be\label{inequality} T S\geq -\int d\Omega_{(D-2)}\int_{BH} d r\,\sqrt{-g}\,\big( \mathcal{L}_{grav}-2\rho\big) \,.\ee

In addition, we point out that for any gravitational object defined in the bulk of WDW patch
\be \mathcal{A}\equiv \int_{WDW}d^D x\,\sqrt{-g}\,\mathcal{F} \,,\ee
its growth rate at late times is given by
\be\label{identity} \fft{d\mathcal{A}}{dt}=\int d\Omega_{(D-2)}\int_{BH} dr\, \sqrt{-g}\,\mathcal{F} \,.\ee
The identity is proved in Appendix \ref{CAHDG} and we refer the readers to there for details.

Based on the above results, we propose that the complexity for a holographic state is given by
\be\label{proposal1} \mathcal{C}=-\fft{\alpha}{\pi\hbar}\int_{WDW}d^D x\sqrt{-g}\,\big(\mathcal{L}_{grav}-2\rho\big) \,,\ee
where the overall constant $\alpha$ has not been fixed by above considerations. As previously, we shall take it to be $\alpha=\ft{2(D-2)}{D-1}$. From the identity (\ref{identity}) and the relation (\ref{inequality}), one has at late times
\bea \fft{d\mathcal{C}}{dt}= -\fft{\alpha}{\pi\hbar}\int d\Omega_{(D-2)}\int_{BH} d r\,\sqrt{-g}\,\big( \mathcal{L}_{grav}-2\rho\big)\leq\fft{\alpha T S}{\pi\hbar }\,.
\eea
We will show this for a variety of examples in the subsequent subsections. In spite of that for our new proposal, the complexity growth rate at late times is not always equal to $\alpha T S/\pi\hbar$, it is still of great interests for our current purpose since we are trying to search such an object whose growth rate is bounded above by the energy alone for the state under considerations.

\subsection{Static black holes}
For static black holes, evaluating the complexity growth at late times yields
\be\label{latecp} \fft{d\mathcal{C}}{dt}\Big|_{late}=\fft{\alpha}{\pi\hbar}\int d\Omega_{(D-2)}\Big(-\sqrt{-g}\,\mathcal{Q}^{rt}_{grav} \Big)\Big|^{r_h}_{\epsilon}    \,,\ee
where it was understood that $\epsilon\rightarrow 0$ for black holes without an inner horizon. Whenever the singularity gives a vanishing contribution, it gives rise to $d\mathcal{C}/dt=\alpha T S/\pi\hbar$. However, without a well developed quantum gravity theory, we shall examine the result more carefully in the classical limit.

We focus on Einstein's gravity which has
\be \mathcal{Q}^{\mu\nu}_{grav}=\fft{1}{16\pi G}\Big(-2\nabla^{[\mu} \xi^{\nu]} \Big) \,.\ee
A straightforward calculation gives
\be \fft{d\mathcal{C}}{dt}\Big|_{late}=\fft{\alpha\, \omega_{D-2}}{16\pi^2\hbar G}\,\fft{h'(r)}{w(r)}\,r^{D-2}\Big|^{r_h}_{\epsilon} \,,\ee
where $w(r)=\sqrt{h(r)/f(r)}$. In the black hole interior $w(r)>0\,,h(r)<0$. As a consequence, close to the singularity, the metric functions behave as to leading order
\be w(r)=c\,r^\gamma+\cdots\,,\quad h(r)=-\fft{\mu}{r^\delta}+\cdots \,,\ee
where $c\,,\mu$ are positive constants. It follows that for $\delta<D-\gamma-3$, the contribution near the singularity vanishes whilst for $\delta=D-\gamma-3$, it is negative definite. As for the case $\delta>D-\gamma-3$, the result diverges (with a negative weight) at the singularity. Again, we cannot manage this credibly in the classical limit. Based on these results, we argue that at late times $d\mathcal{C}/dt\leq \alpha T S/\pi\hbar$ in the classical limit.

For black holes with an inner horizon, one should take $r_h=r_+\,,\epsilon=r_-$ in Eq.(\ref{latecp}). One finds
\be \fft{d\mathcal{C}}{dt}\Big|_{late}=\fft{\alpha}{\pi\hbar}\Big(T_+S_+-T_-S_- \Big)\,,\ee
where $T_-$ is the negative temperature defined on the inner horizon.

\subsection{Rotating BTZ black hole}
We would like to extend our discussions to the rotating BTZ black hole
\bea\label{btz}
&&ds^2=-f(r)dt^2+\fft{dr^2}{f(r)}+r^2\Big(d\theta-\fft{4GJ}{r^2}\,dt \Big)^2 \,,\nn\\
&&f(r)=r^2\ell^{-2}-8GM+\fft{16G^2 J^2}{r^2}\,,
\eea
where $M$ and $J$ are the black hole mass and angular momentum respectively. The solution in general has two horizons. One has
\be \fft{r_+^2+r_-^2}{\ell^2}=8GM\,,\quad \fft{2r_+r_-}{\ell^2}=\fft{8GJ}{\ell} \,.\ee
Clearly, the angular momentum is bounded above $J\leq M\ell$. The bound will be saturated when the solution becomes extremal. The temperatures, entropies and angular velocities on the horizons are given by
\be T_\pm=\fft{r_\pm^2-r_\mp^2}{2\pi\ell^2 r_\pm }\,,\quad S_\pm=\fft{\pi r_\pm}{2G}\,,\quad \Omega_\pm=\fft{r_\mp}{r_\pm \ell} \,.\ee
 It turns out that though the solution has rotations, our formula (\ref{bulk}) continues hold. In addition, there exists a global scaling symmetry
\be r\rightarrow \lambda r\,,\quad (t\,,\theta)\rightarrow \lambda^{-1}(t\,,\theta)\,,\quad f(r)\rightarrow \lambda^2 f(r) \,,\ee
which implies that
\be M\rightarrow \lambda^2 M\,,\quad T\rightarrow \lambda T\,,\quad S\rightarrow \lambda S\,,\quad\Omega\rightarrow \Omega\,,\quad J\rightarrow \lambda^2 J \,.\ee
This leads to a generalised Smarr formula
\be M-\Omega J=\ft 12 T S \,,\ee
which holds on both horizons. Therefore, at late times, the complexity growth for a rotating BTZ black hole is given by
\bea
\fft{d\mathcal{C}}{dt}&=&\fft{1}{\pi\hbar}\big(T_+S_+-T_-S_- \big) \nn\\
&=&\fft{2}{\pi\hbar}\Big[ \big( M-\Omega J\big)_+-\big(M-\Omega J \big)_- \Big]\,.
\eea
The angular momentum plays a same role as the electric charge in Eq.(\ref{cpinner}). Without further discussions, we present the result for a rotating charged BTZ black hole in high temperature limit
\be\fft{d\mathcal{C}}{dt}
=\fft{2}{\pi\hbar}\Big[ \big( M-\mu Q-\Omega J\big)_+-\big(M-\mu Q-\Omega J \big)_- \Big]_{T\rightarrow \infty}\,. \ee

\subsection{Kerr-AdS black hole}
We continue studying the complexity growth rate at late times for a Kerr-AdS black hole
\be\label{kerr4} ds^2=-\fft{\Delta_r}{\rho^2}\Big(dt-a\sin{\theta}^2 \fft{d\varphi}{\Xi} \Big)^2+\rho^2\Big(\fft{dr^2}{\Delta_r}+\fft{d\theta^2}{\Delta_\theta} \Big)+\fft{\Delta_\theta \sin{\theta}^2}{\rho^2}\Big(a dt-(r^2+a^2) \fft{d\varphi}{\Xi} \Big)^2 \,,\ee
where $a=J/M$ is the ratio of angular momentum to the black hole mass and
\bea
&&\Xi=1-g^2 a^2\,,\nn\\
&&\rho^2=r^2+a^2 \cos{\theta}^2\,,\nn\\
&&\Delta_\theta=1-g^2a^2\cos{\theta}^2\,,\nn\\
&&\Delta_r=(r^2+a^2)(1+g^2r^2)-2GM \Xi^2 r\,.
\eea
The horizons are defined by $\Delta_r(r_\pm)=0$. The angular velocities, entropies and temperatures are given by
\bea
&&\Omega_\pm= \fft{a(g^2r_\pm^2+1)}{r_\pm^2+a^2}\,,\quad S_\pm=\fft{\pi (r_\pm^2+a^2)}{G\Xi}\,,\nn\\
&&T_\pm=\fft{3g^2r_\pm^4+(1+g^2a^2)r^2_\pm-a^2}{4\pi r_\pm (r^2_\pm+a^2)}\,.
\eea
It follows that at high temperatures, one has to leading order
\be M-\Omega J\simeq \ft{2}{3}\,T S \,.\ee
Again the angular momentum plays a same role as the electric charge. However, now our formula (\ref{bulk}) for the Noether charge is invalid. Instead, to derive the complexity growth, one should directly calculate the bulk integral (\ref{proposal1}). One finds
\be \fft{d\mathcal{C}}{dt}=\fft{2g^2r(r^2+a^2) }{3\pi\hbar G \Xi}\Big|^{r_+}_{r_-} \,.\ee
It follows that in high temperature limit
\bea \fft{d\mathcal{C}}{dt}&\simeq& \fft{4}{3\pi\hbar}\Big[\big(M-\Omega J \big)_+ -\big( M-\Omega J \big)_- \Big] \nn\\
&<&\fft{2}{\pi\hbar}\Big[\big(M-\Omega J \big)_+ -\big( M-\Omega J \big)_- \Big]\,.
\eea

Based on all the results above and considering the relevance to the Lloyd bound, we conjecture that for our new proposal, the complexity growth for general holographic states obeys the upper bound $d\mathcal{C}/dt\leq \alpha T S/\pi\hbar$ or $d\mathcal{C}/dt\leq \alpha \big( T_+ S_+-T_-S_- \big)/\pi\hbar$ for black holes with an inner horizon.

\section{The time dependence of complexity and the upper bound}
\begin{figure}[htbp]
  \centering
  \includegraphics[width=3.5in]{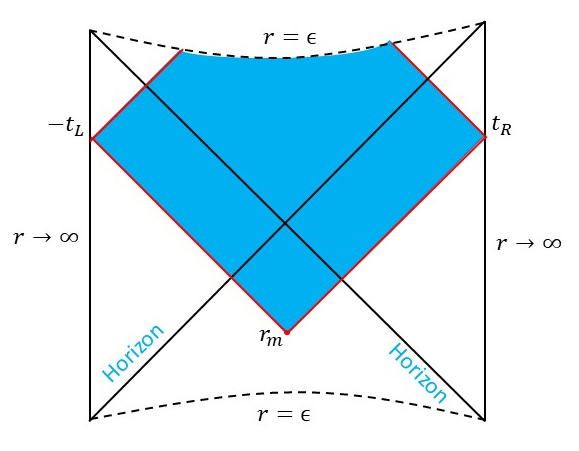}
  	\caption{The Wheeler-DeWitt (WDW) patch of a neutral two-sided AdS black hole. It moves forward in a symmetric way $t_L=t_R=t/2$. The dotted lines $r=\epsilon$ denote the locus of spacelike singularities, where the WDW patch terminates. }
\label{tfd} \end{figure}

To test our conjecture, we shall extract the time dependence of complexity at finite times for stationary AdS black holes. The Fig.\ref{tfd} shows the WDW patch of a neutral two-sided AdS black hole. There are two boundary times $t_L\,,t_R$ but the complexity according to the symmetry of the problem, is only a function of the combination $t\equiv t_L+t_R$.

By following the prescriptions in \cite{Carmi:2017jqz}, one finds
 \bea\label{timebulk}
\fft{d\mathcal{C}}{dt}&=&-\fft{\alpha}{\pi\hbar}\int d\Omega_{D-2}\int_{\epsilon}^{r_m}\mathrm{d}r\,\sqrt{- g}\,\big( \mathcal{L}_{grav}-2\rho\big)\,.
\eea
Notice that the time dependence of complexity is implicitly contained in the evolution of the position $r_m$, which is determined by
\be\label{positionrm} t\equiv t_R+t_L=-2r^*(r_m) \,,\ee
where the tortoise coordinate is defined as $r^*(r)=-\int^\infty_r \mathrm{d}r/w(r)f(r)$. However, instead of numerically studying the complexity growth in a case-by-case basis, we find that under the condition
 \be
 \mathcal{L}_{grav}-2\rho\leq 0\,,\quad  \epsilon < r\leq r_h\label{condition1}\,,
 \ee
 the states will obey the conjectured upper bound. Interestingly, the condition looks very much like an energy condition. Indeed, for black holes in Einstein's gravity, it is already guaranteed by known energy conditions for ordinary matter fields, as will be shown later. Another weaker but independent condition is
 \bea &&\big(\mathcal{L}_{grav}-2\rho\big)_{r=r_h}<0\,,\nn\\
  && \big(\mathcal{L}_{grav}-2\rho)_{\epsilon<r<r_h}=0\quad has\,\, one\,\, real\,\, root\,\,at\,\,most\,.\label{condition2}
 \eea

To derive the above conditions, we first notice that there is a critical time $t_c=-2r^*(\epsilon)$, at which the position $r_m$ lifts off of the past singularity \cite{Carmi:2017jqz}. When $t\leq t_c$, $r_m=\epsilon$, the complexity remains a constant and hence $d\mathcal{C}/dt=0$. However, when $t>t_c$, the position $r_m$ begins to grow with time. Its growth rate is given by
\be \fft{dr_m}{dt}=-\ft 12 w(r_m)f(r_m)\geq 0 \,.\ee
Thus, when $t>t_c$, $r_m$ monotonically increases and approaches the event horizon $r_h$ from below at late times. Taking one more derivative with respect to $t$ for Eq.(\ref{timebulk}), we find
\be \fft{d^2\mathcal{C}}{dt^2}=-\fft{\omega_{D-2}}{\pi\hbar}\Big[\sqrt{-\bar g}\big(\mathcal{L}_{grav}-2\rho\big)\Big]_{r=r_m}\,\dot{r}_m \geq 0\,,\ee
under the condition (\ref{condition1}).
 Therefore, the rate of change of complexity is a monotone increasing function of time and hence $d\mathcal{C}/dt\leq \alpha TS/\pi\hbar$.

On the other hand, even if the condition (\ref{condition1}) no longer holds, one can instead derive the alternative condition (\ref{condition2}) by examining the behavior of complexity at late times. One has
\be r_m=r_h-c_m\, e^{-2\pi T t}+\cdots \,,\ee
where $c_m$ is a positive constant. A straightforward calculation gives
\be \fft{d\mathcal{C}}{dt}=\fft{\alpha T S}{\pi\hbar}+\fft{\omega_{D-2}}{16\pi G}\Big[\sqrt{-\bar g}\big(\mathcal{L}_{grav}-2\rho \big)\Big]_{r=r_h}\,c_m\, e^{-2\pi T t}+\cdots \,.\ee
Thus, if $\big( \mathcal{L}_{grav}-2\rho\big)_{r=r_h}<0$, the late time rate of change of complexity will be always approached from below. Moreover, if the equation $\big(\mathcal{L}_{grav}-2\rho \big)_{\epsilon<r<r_h}=0$ does not have a real root, then the condition (\ref{condition1}) will be satisfied. Instead, if $\big(\mathcal{L}_{grav}-2\rho \big)_{\epsilon<r<r_h}=0$ has one real root, $d\mathcal{C}$/dt will have a local extrema when the position $r_m$ meets the root. However, it cannot be a local maxima since the late time rate of change of complexity is approached from below. Therefore, under the condition (\ref{condition2}) the complexity growth again obeys the bound $d\mathcal{C}/dt\leq \alpha TS/\pi\hbar$.

 For black holes with an inner horizon, one can follow the above discussions and derive similar conditions
  \be
 \mathcal{L}_{grav}-2\rho\leq 0\,,\quad  r_- \leq r\leq r_+\label{condition31}\,,
 \ee
 or
 \bea &&\big(\mathcal{L}_{grav}-2\rho\big)_{r=r_\pm}<0\,,\nn\\
 && \big(\mathcal{L}_{grav}-2\rho)_{r_-<r<r_+}=0\quad has\,\, one\,\, real\,\, root\,\,at\,\,most\,.\label{condition32}
 \eea
 We refer the readers to Appendix \ref{appcondition} for details.

 \subsection{Einstein-Maxwell-Dilaton black holes}
 By using these conditions, we would like to analytically test our conjecture for a variety of stationary AdS black holes.

First of all, for Schwarzschild black holes, one has $\rho=\Lambda=-\ft 12(D-1)(D-2)\ell^{-2}$ and
\be \mathcal{L}_{grav}-2\rho=R-2\Lambda=-2(D-1)\ell^{-2}<0\,,\ee
exactly satisfying the condition (\ref{condition1}). Likewise, for the Kerr-AdS black hole (\ref{kerr4}), one finds
 \be\mathcal{L}_{grav}-2\rho=-\fft{6\ell^{-2}}{\Sigma}<0\,. \ee
 In the presence of matter fields, by using Einstein's equations $R_{\mu\nu}-\fft 12 g_{\mu\nu}R=T_{\mu\nu}$, one finds
\be \mathcal{L}_{grav}-2\rho=-\ft{2}{D-2}\,\Delta \,,\quad \Delta=(D-3)\rho+\sum_{\alpha} p_\alpha\,,\ee
where $p_\alpha=T^\alpha_{\,\,\alpha}$, $\alpha=r\,,x_i$ are the principal pressures of matter fields. The condition (\ref{condition1}) turns out to be equivalent to $\Delta\geq 0$. However, this is already guaranteed by weak and strong energy conditions. Thus, for ordinary matter fields (without ghost like modes), the condition (\ref{condition1}) is easily satisfied in AdS spacetimes.

For example, for hairy black holes in Einstein-Scalar gravity
\be  \mathcal{L}=R-\ft 12\big( \partial \phi\big)^2-V(\phi) \,,\ee
one has
\be \rho=-p_i=\ft 12 V(\phi)+\ft 14 f\phi'^2\,,\qquad p_r=p_i+\ft12 f\phi'^2 \,.\ee
It follows that $\Delta=-V(\phi)\geq 0$, which is a least condition to guarantee the asymptotic structure of AdS spacetimes.

For electrically charged black holes in Einstein-Maxwell-Dilaton theories
\be  \mathcal{L}=R-\ft 12\big( \partial \phi\big)^2-V(\phi)-Z(\phi)F^2\,,\quad A=a(r)dt\,,\ee
one has
\bea
&& \rho=\ft 12 V(\phi)+\ft 14 f\phi'^2+Z(\phi)a'^2/w^2\,,\nn\\
&&p_r=-\ft 12 V(\phi)+\ft 14 f\phi'^2-Z(\phi)a'^2/w^2\,,\nn\\
&&p_i=-\ft 12 V(\phi)-\ft 14 f\phi'^2+Z(\phi)a'^2/w^2\,.
\eea
Again, $\Delta=-V(\phi)+2(D-3)Z(\phi)a'^2/w^{2}\geq 0$. We conclude that the condition (\ref{condition1}) is valid to a lot of physically interesting solutions, including supersymmetric black holes.

\subsection{Gauss-Bonnet black holes}
We continue testing our conjecture for black holes with stringy corrections. An interesting example is the Gauss-Bonnet black holes (we focus on the planar case so that at late times $d\mathcal{C}/dt=2M/\pi\hbar$)
\bea\label{GBBH}
&&ds^2=-f(r)dt^2+\fft{dr^2}{f(r)}+\sum_{i=1}^{D-2}r^2 dx^i dx^i\,,\quad f(r)=\ft{r^2}{2\lambda\ell^2}\Big(1-\varphi(r) \Big)\,,
\eea
where the coupling constant $\lambda$ should be positive according to string theory and the function $\varphi(r)$ is given by
\be
\varphi(r)=\sqrt{1-4\lambda\Big(1-\ft{r_h^{D-1}}{r^{D-1}} \Big)}\,.
\ee
The event horizon is defined as $\varphi(r_h)=1$ and the black hole interior corresponds to $\varphi\geq 1$. It was established in \cite{Brigante:2007nu,Camanho:2009vw} that the Gauss-Bonnet coupling was strongly constrained by microcausality of the theories (or equivalently positive energy fluxes in scattering process). The allowed region for the coupling constant is \cite{Brigante:2007nu,Camanho:2009vw}
\be\label{lambda} 0< \lambda\leq \ft{(D-3)(D-4)(D^2-3D+8)}{4(D^2-5D+10)^2} \,.\ee
This is the parameter space that we should study carefully. Evaluating the Lagrangian density yields
\bea
\mathcal{L}_{EGB}&=&R+(D-1)(D-2)\ell^{-2}+\ft{\lambda\,\ell^{2}}{(D-3)(D-4)}\big(R^2-4R^2_{\mu\nu}+R^2_{\mu\nu\lambda\rho} \big) \,,\nn\\
&=&-\ft{D-1}{4(D-4)\lambda\ell^2\varphi^3}\,\Big[(D+1)\,\varphi^4-4\Big(8\lambda+D(1-4\lambda) \Big)\,\varphi^3\nn\\
&&\qquad\qquad\qquad\quad+2(2D-1)(1-4\lambda)\,\varphi^2-(D-1)(1-4\lambda)^2\Big]\,.
\eea
It follows that in the allowed region for the coupling constant, one has on the horizon
\be \mathcal{L}_{EGB}(r=r_h)=-\ft{4(D-1)^2}{D-4}\Big(\ft{D-4}{2(D-1)} -\lambda\Big)<0 \,,\ee
 and near the singularity
 \be \mathcal{L}_{EGB}|_{r\rightarrow \epsilon}\sim -\ft{(D^2-1)\varphi}{4(D-4)\lambda\ell^2}\rightarrow -\infty\,.\ee
 However, for a certain range of parameters $0<\lambda<\ft{D-2}{6(D-1)}$, the Lagrangian density will have a dangerously local maxima at $\varphi_{max}=\sqrt{\ft{3(D-1)(1-4\lambda)}{D+1}}$, which might be positive definite. It turns out that there is a critical Gauss-Bonnet coupling, given by
 \begin{figure}[ht]
\centering
\includegraphics[width=140pt]{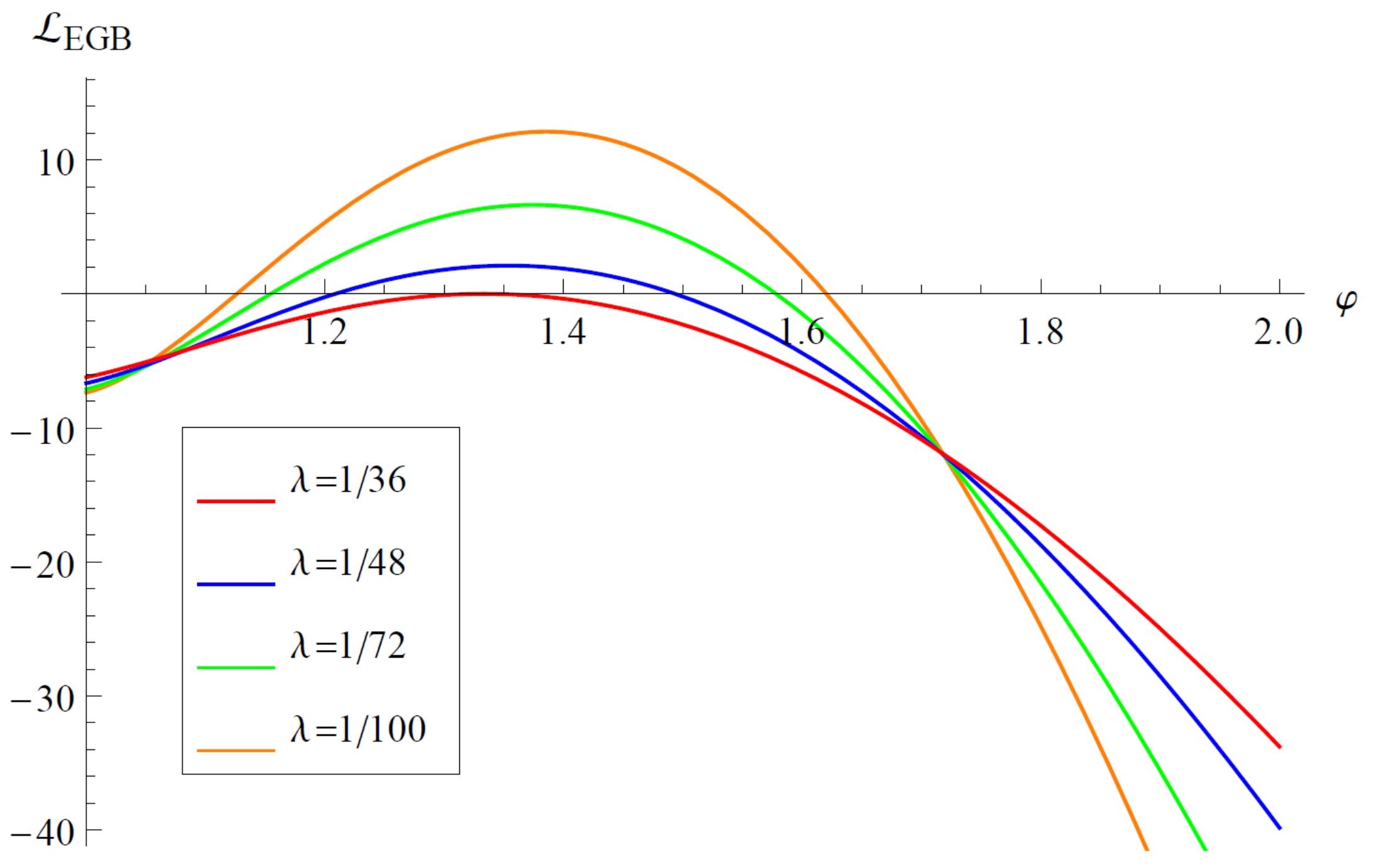}
\includegraphics[width=140pt]{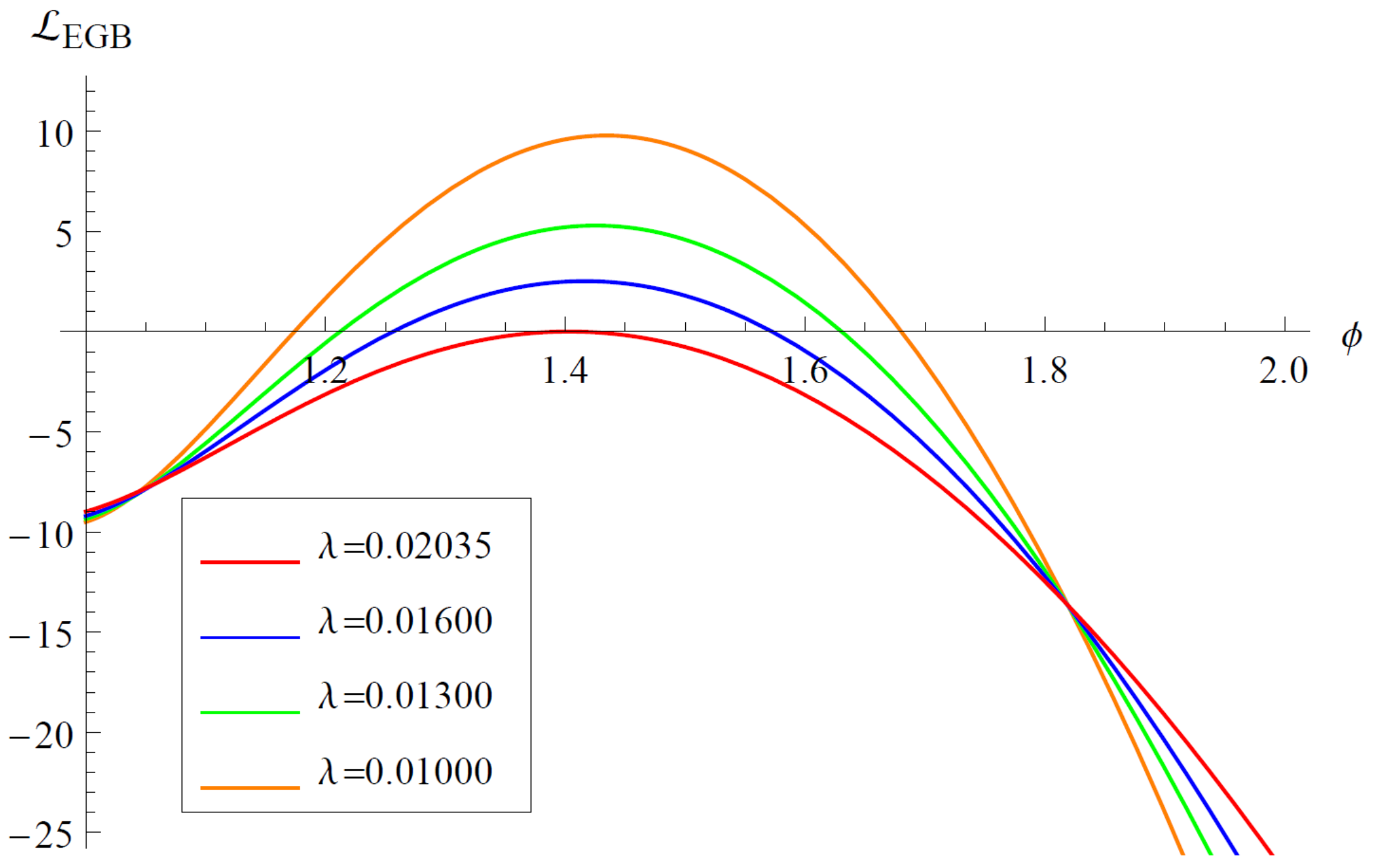}
\includegraphics[width=140pt]{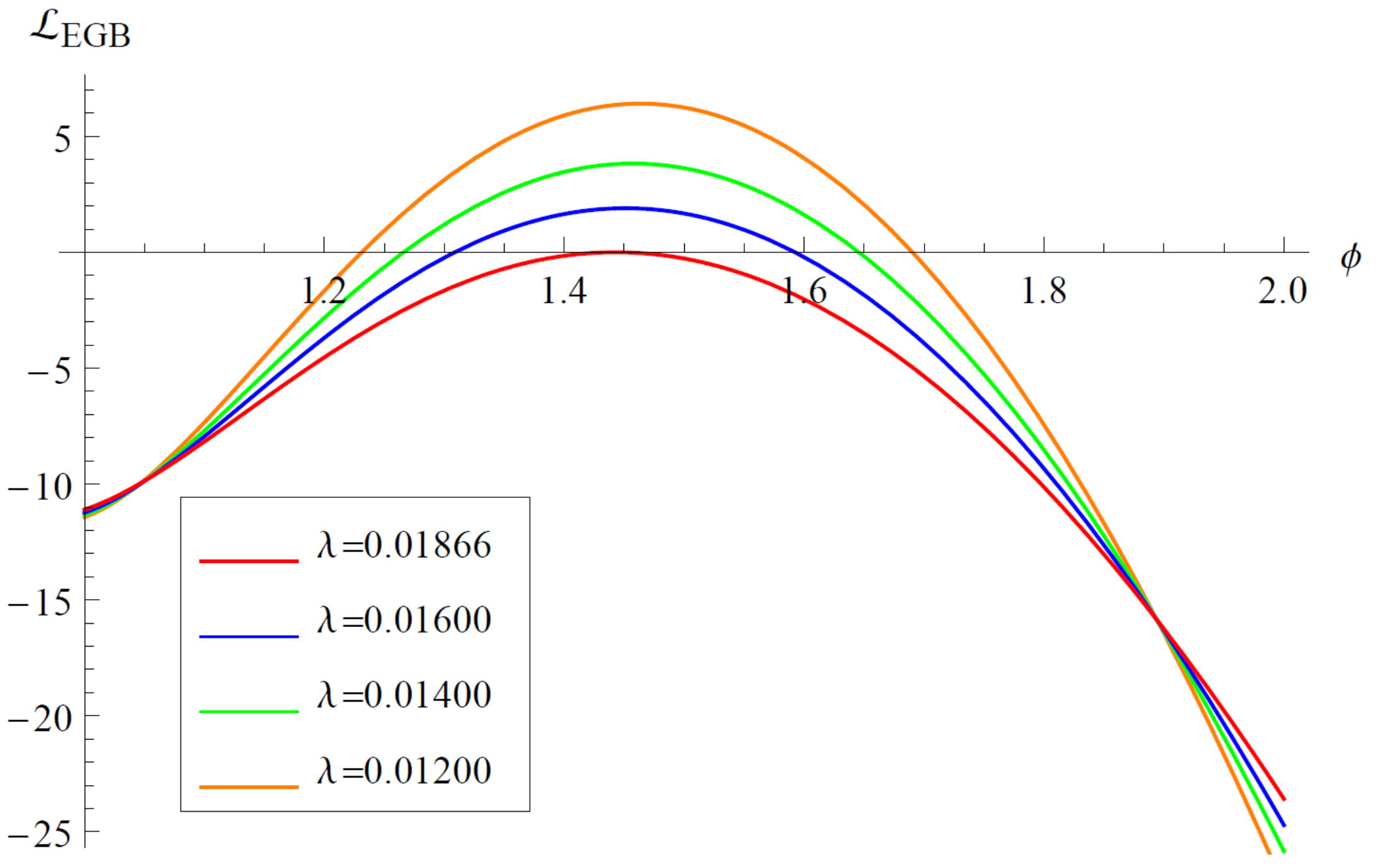}
\caption{{\it The plots for $\mathcal{L}_{EGB}$ in $D=5\,,6\,,7$ dimensions as a function of $\varphi$. For $0<\lambda<\lambda_c$, the Lagrangian density always has two real roots in the black hole interior and a positive local maxima between them.  }}
\label{GB}\end{figure}
\be
\lambda_c=\ft{\Big(29D^3-147D^2+132D-16-(5D-4)\sqrt{(D+1)\big(25D^3-231D^2+624D-416\big)} \Big) }{216(D-1)(D-2)^2}\,,\ee
beyond which $\lambda\geq \lambda_c$ one safely has $\mathcal{L}_{EGB}\Big|_{\epsilon<r<r_h}\leq 0$ and hence the condition (\ref{condition1}) is satisfied. For example, for $D=5$, $\lambda_c=1/36\simeq 0.02778$, for $D=6$, $\lambda_c\simeq 0.02035$, for $D=7$, $\lambda_c\simeq 0.01866$ and in the large $D$ limit, $\lambda_c\rightarrow 1/54\simeq 0.01852$.

\begin{figure}[ht]
\centering
\includegraphics[width=230pt]{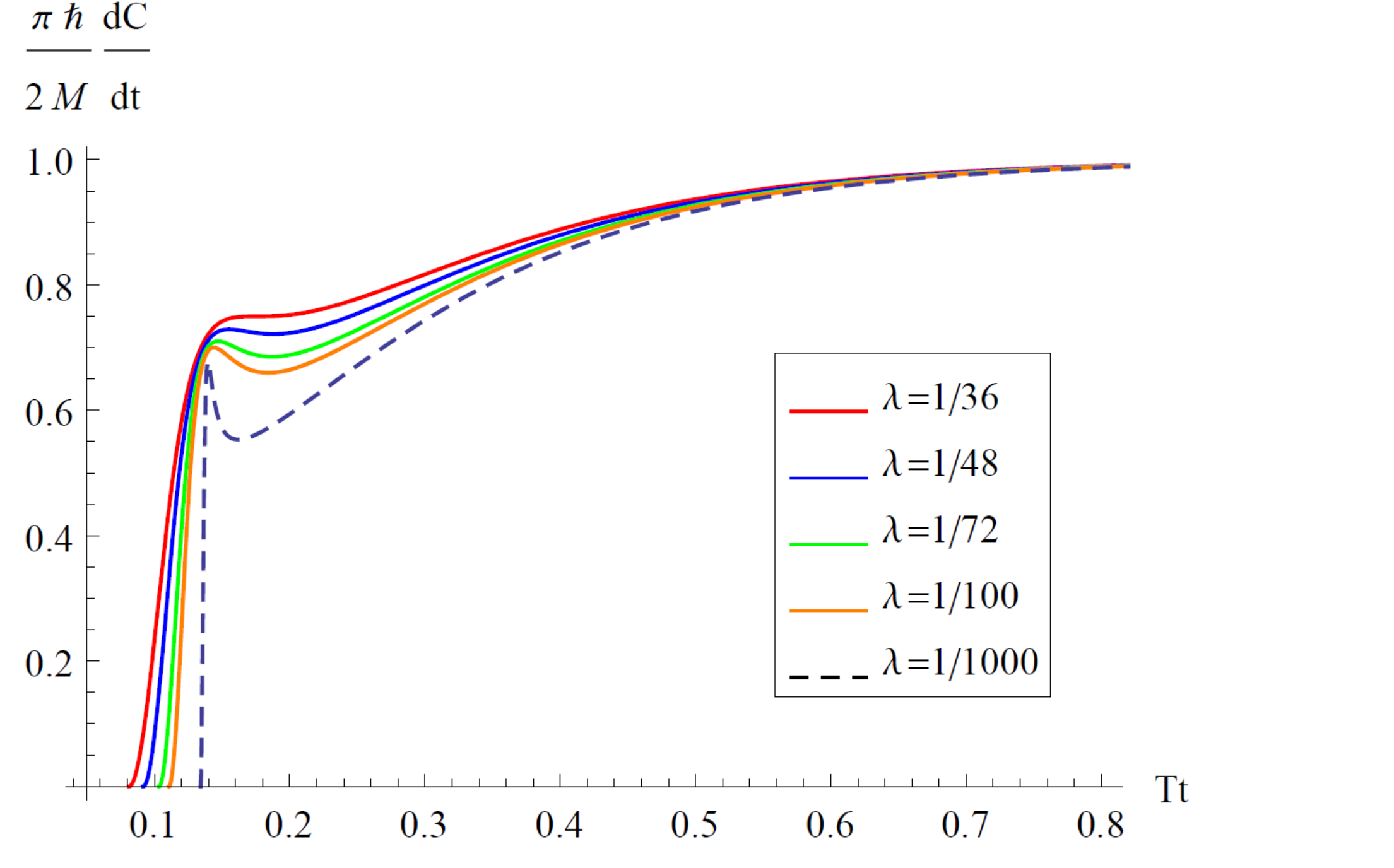}
\includegraphics[width=210pt]{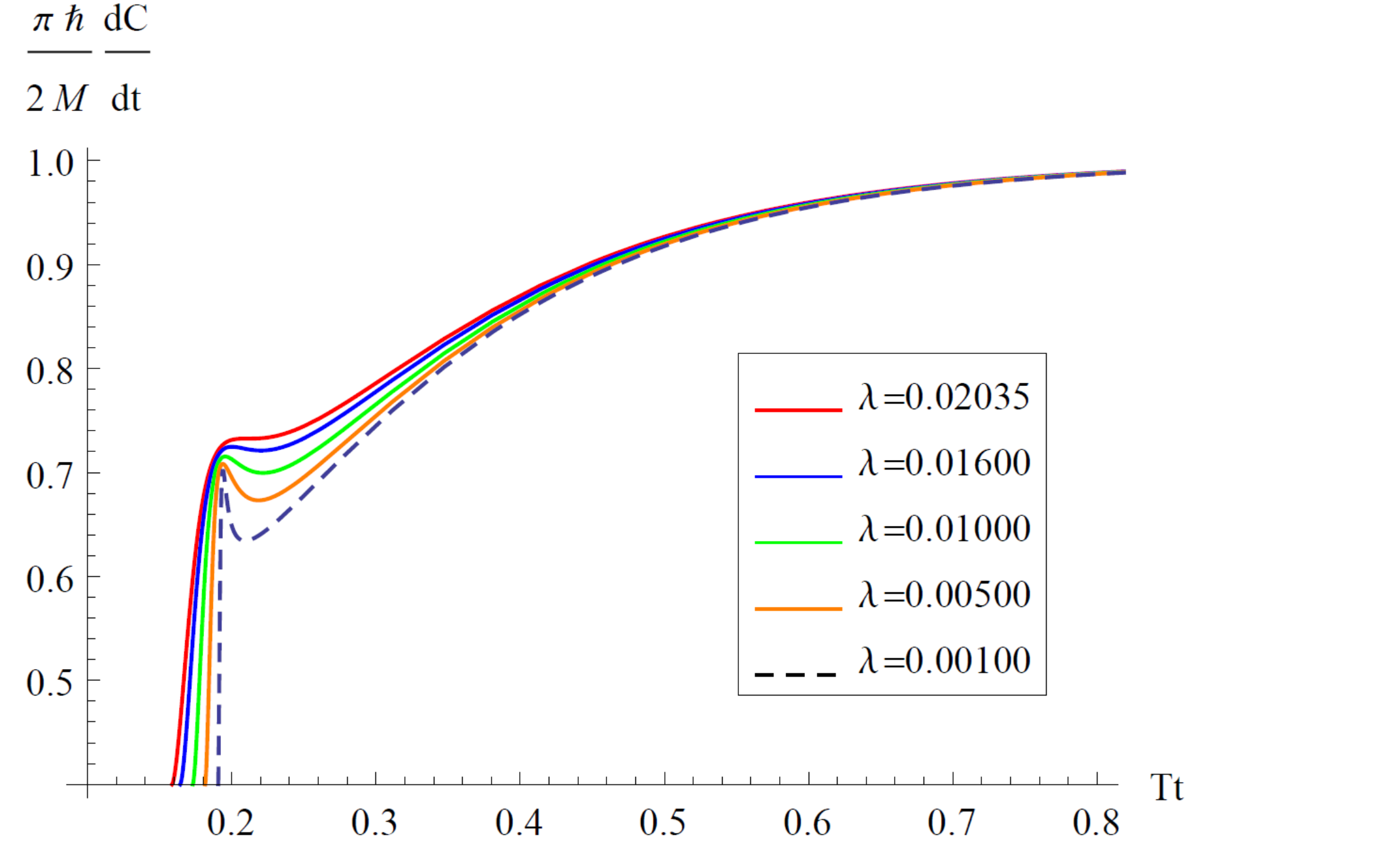}
\includegraphics[width=210pt]{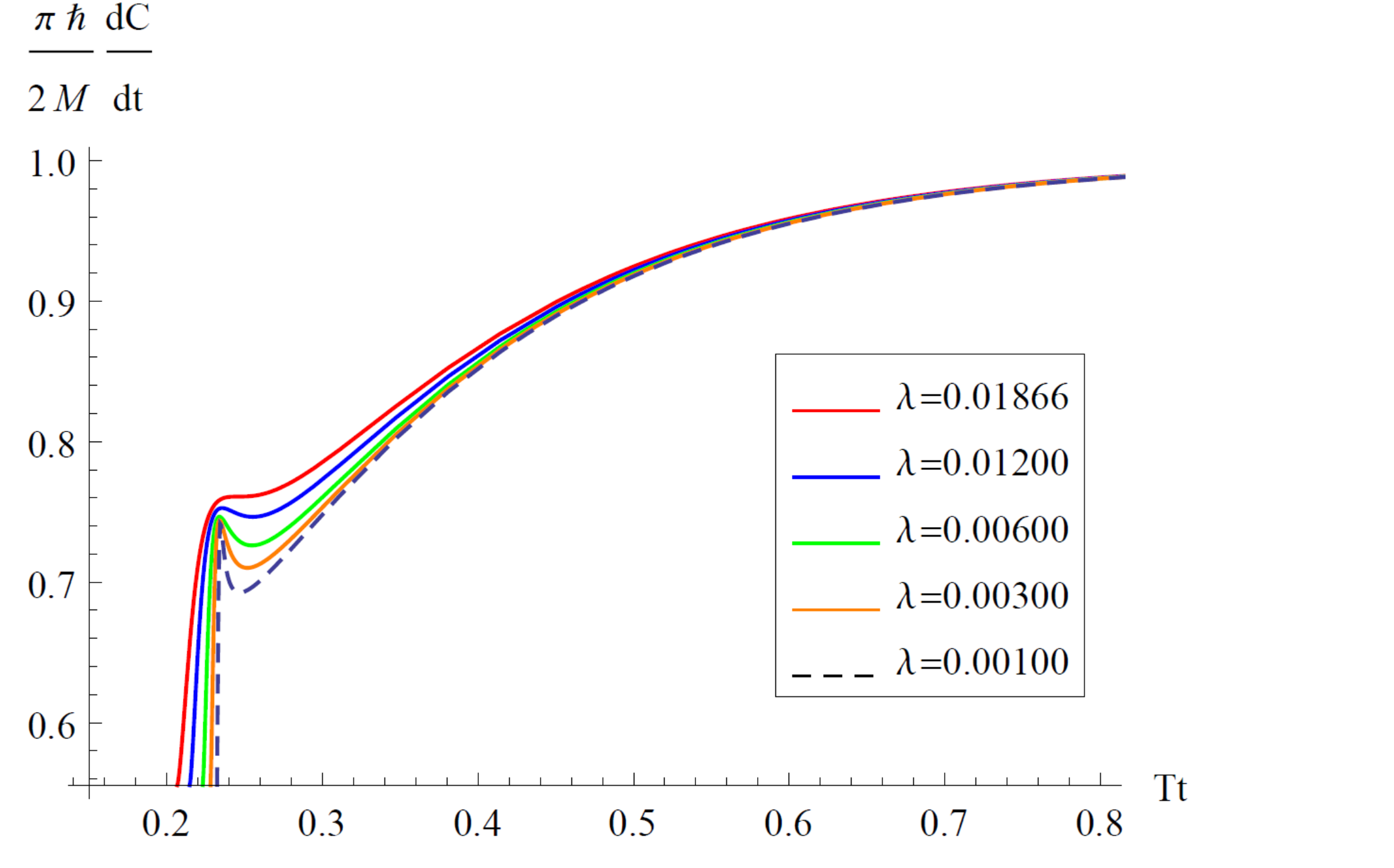}
\caption{{\it The rate of change of complexity for Gauss-Bonnet black holes in $D=5\,,6\,,7$ dimensions with $0<\lambda<\lambda_c$.  }}
\label{GBcp}\end{figure}

However, when $0<\lambda<\lambda_c$, the Lagrangian density will be positive definite in a certain region of black holes, see Fig.\ref{GB}. Thus, this case is not covered by our conditions (\ref{condition1}) and (\ref{condition2}). Instead, we have to adopt numerical approach to study the evolution of complexity at finite times. Our numerical results (see Fig.\ref{GBcp}) suggest that the complexity growth still obeys the Lloyd bound for $0<\lambda<\lambda_c$ for Gauss-Bonnet black holes in any number of dimensions.

In fact, since $\lambda$ is small, we can examine this by means of half-analytical approach as well.
At early times, the complexity growth behaves as
\be \fft{d\mathcal{C}}{dt}\Big|_{early}=\fft{2M}{\pi\hbar}\Big[\ft{(D-2)(D-5)}{(D-1)(D-4)}+O\Big(T(t-t_c)\Big)^{\fft{D-1}{D-3}}\, \Big] \,,\ee
and it will approach a local maxima when the corner $r_m$ of the WDW patch meets the larger root of the equation $\mathcal{L}_{EGB}=0$. Our goal is to calculate the local maximal growth rate in terms of a small $\lambda$ series and compare it with the growth rate at late times. By simple calculations, we deduce
\be
\fft{\pi\hbar}{2M}\fft{d\mathcal{C}_{local\,max}}{dt}=
\left\{
\begin{array}{ll}
0.675445+2.370771\,\lambda+8.58415\,\lambda^2+45.2119\,\lambda^3+\cdots\,,\quad D=5\,,\\
0.701613+1.294221\,\lambda+7.52887\,\lambda^2+83.4051\,\lambda^3+\cdots\,,\quad D=6\,,\\
0.741077+0.885156\,\lambda+6.14805\,\lambda^2+81.0730\,\lambda^3+\cdots\,,\quad D=7\,,\\
0.773856+0.667725\,\lambda+5.08560\,\lambda^2+72.2716\,\lambda^3+\cdots\,,\quad D=8\,,\\
0.800000+0.533333\,\lambda+4.29630\,\lambda^2+63.5193\,\lambda^3+\cdots\,,\quad D=9\,,\\
0.821007+0.442464\,\lambda+3.70058\,\lambda^2+56.0147\,\lambda^3+\cdots\,,\quad D=10\,,\\
0.838149+0.377174\,\lambda+3.24017\,\lambda^2+49.7928\,\lambda^3+\cdots\,,\quad D=11\,.
\end{array}\right.
\ee
\begin{figure}[ht]
\centering
\includegraphics[width=250pt]{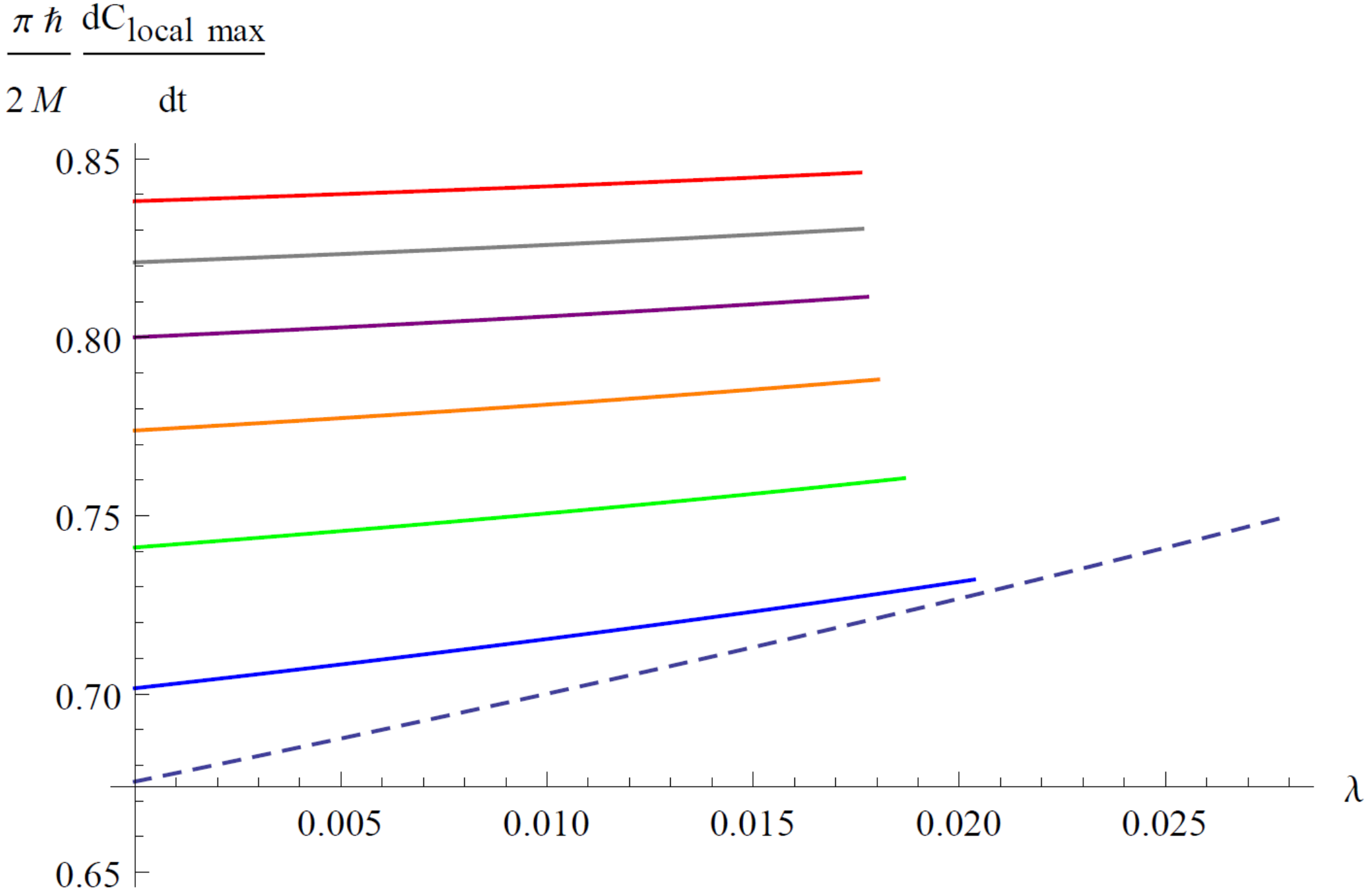}
\caption{{\it The local maxima of complexity growth for $0<\lambda<\lambda_c$. From bottom to top, $D=5$ (dashed), $D=6$ (blue), $D=7$ (green), $D=8$ (orange), $D=9$ (purple), $D=10$ (gray), $D=11$ (red).  }}
\label{GBcp2}\end{figure}
Notice that the leading order is always less than unity but it increases with spacetime dimensions. In fact, by plotting the local maximal growth rate as a function of $\lambda$ (see Fig.\ref{GBcp2}), we find that in each dimension, it is always smaller than the Lloyd bound for $0<\lambda<\lambda_c$. However, for a given $\lambda$, the ratio $\fft{\pi \hbar}{2M}\fft{d\mathcal{C}}{dt}$ increases with spacetime dimensions. Thus, we would like to investigate the result in the large $D$ limit. A straightforward derivation gives
\be \fft{\pi\hbar}{2M}\fft{d\mathcal{C}_{local\,max}}{dt}=1-\fft{1.656854(0.707107-\lambda)}{D}+\cdots\,. \ee
Interestingly, in spite of that the leading order approaches unity, the next-to-leading order is always negative definite.

Bases on these results, we conclude that the complexity growth for the allowed coupling constants obeys the Lloyd bound for Gauss-Bonnet black holes in any number of dimensions.

\subsection{Lovelock black holes}
\begin{figure}[ht]
\centering
\includegraphics[width=270pt]{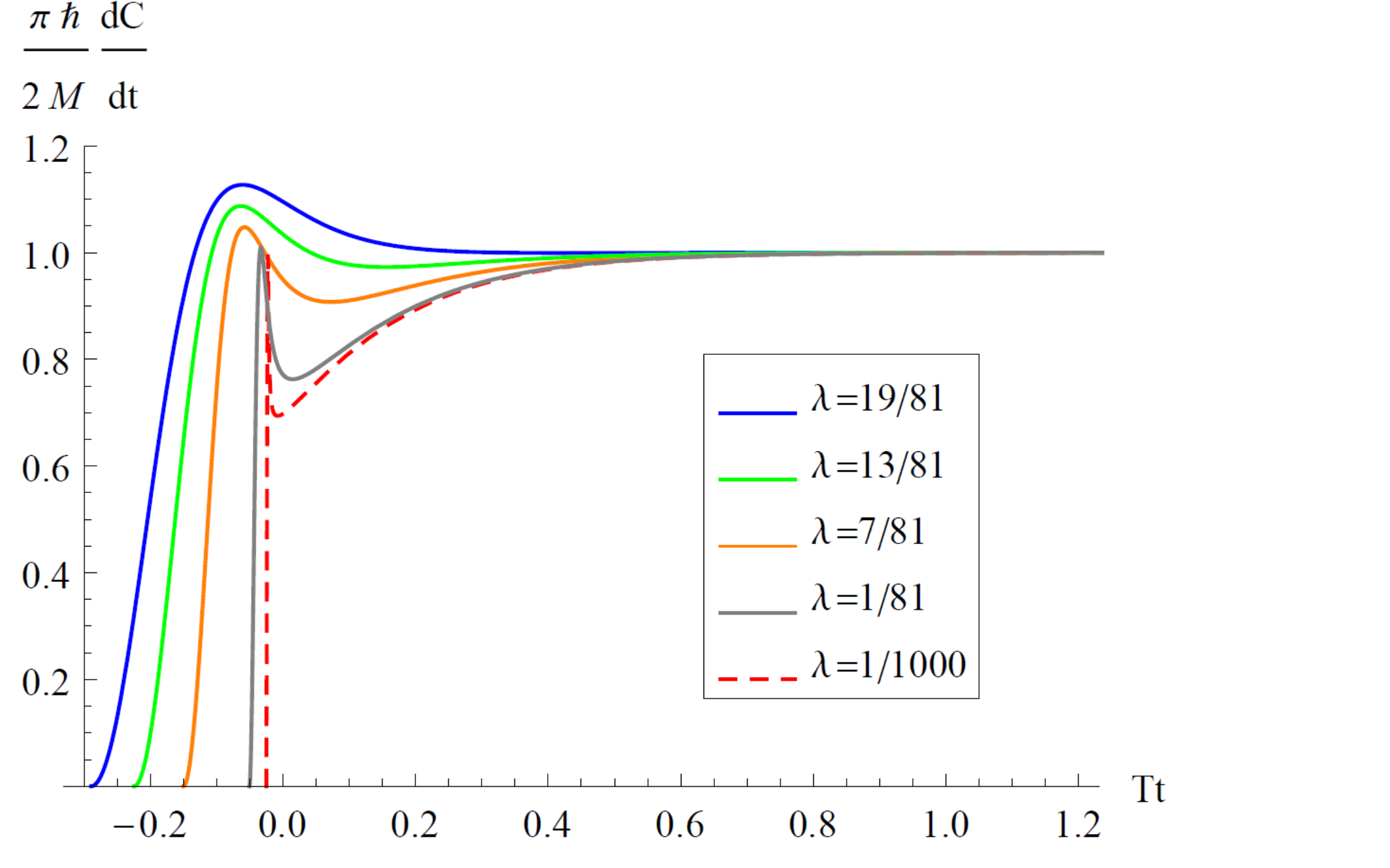}
\caption{{\it The time dependence of complexity for the Lovelock black hole (\ref{LL3BH}). The local maximal growth rate of complexity always exceeds the Lloyd bound for any given $\lambda$ in the allowed region.}}
\label{LL3cp}\end{figure}

To check whether our conjecture can be violated in the presence of stringy corrections, we further study the complexity growth rate for black holes in third order Lovelock gravity. We focus on the $D=7$ dimensional solution. The Lagrangian density reads
\be\mathcal{L}_{tot}=R+30\ell^{-2}+\ft{\lambda\,\ell^{2}}{12}\big(R^2-4R^2_{\mu\nu}+R^2_{\mu\nu\lambda\rho} \big)+\ft{\mu\,\ell^4}{72}\mathcal{L}_{3} \,,\ee
where $\mu$ is the third order gravitational coupling constant and
\bea
\mathcal{L}_3&=&R^3+3R R^{\mu\nu\alpha\beta}R_{\alpha\beta\mu\nu}-12R R^{\mu\nu}R_{\mu\nu}+24R^{\mu\nu\alpha\beta}R_{\mu\alpha}R_{\nu\beta}+16R^{\mu\nu}R_{\nu}^{\,\,\,\alpha}R_{\mu\alpha}\nn\\
&&+24R^{\mu\nu\alpha\beta}R_{\alpha\beta\nu\rho}R_{\mu}^{\,\,\,\rho}
+8R^{\mu\nu}_{\,\,\,\,\,\,\alpha\rho}R^{\alpha\beta}_{\,\,\,\,\,\,\nu\sigma} R^{\rho\sigma}_{\,\,\,\,\,\,\mu\beta}+2R_{\alpha\beta\rho\sigma}R^{\mu\nu\alpha\beta}R^{\rho\sigma}_{\,\,\,\,\,\, \mu\nu}\,.
\eea
The black hole solution can be formally written as
\be\label{LL3BH} ds^2=-f(r)dt^2+\fft{dr^2}{f(r)}+r^2 \sum_{i=1}^{5}dx^i dx^i\,,\quad f(r)=\ft{\lambda r^2}{\mu \ell^2}\Big(1-\varphi(r) \Big)  \,,\ee
where the form of the function $\varphi(r)$ is complicated for a generic $\mu$ (see for example \cite{deBoer:2009gx,Camanho:2009hu}). In the following, we focus on a special case $\mu=\lambda^2$. The function $\varphi(r)$ simplifies to
\be \varphi(r)=\Big[1-3\lambda\Big(1-\ft{r_h^6}{r^6} \Big) \Big]^{1/3} \,.\ee
Again the event horizon is defined as $\varphi(r_h)=1$ and the black hole interior corresponds to $\varphi(r)\geq 1$. However, the new coupling constant $\mu$ strongly effects the causal structure of the theory so that the allowed regime for $\lambda$ is changed \cite{deBoer:2009gx,Camanho:2009hu}. One has
\be 0< \lambda\leq \fft{19}{81}\,. \ee
\begin{figure}[ht]
\centering
\includegraphics[width=250pt]{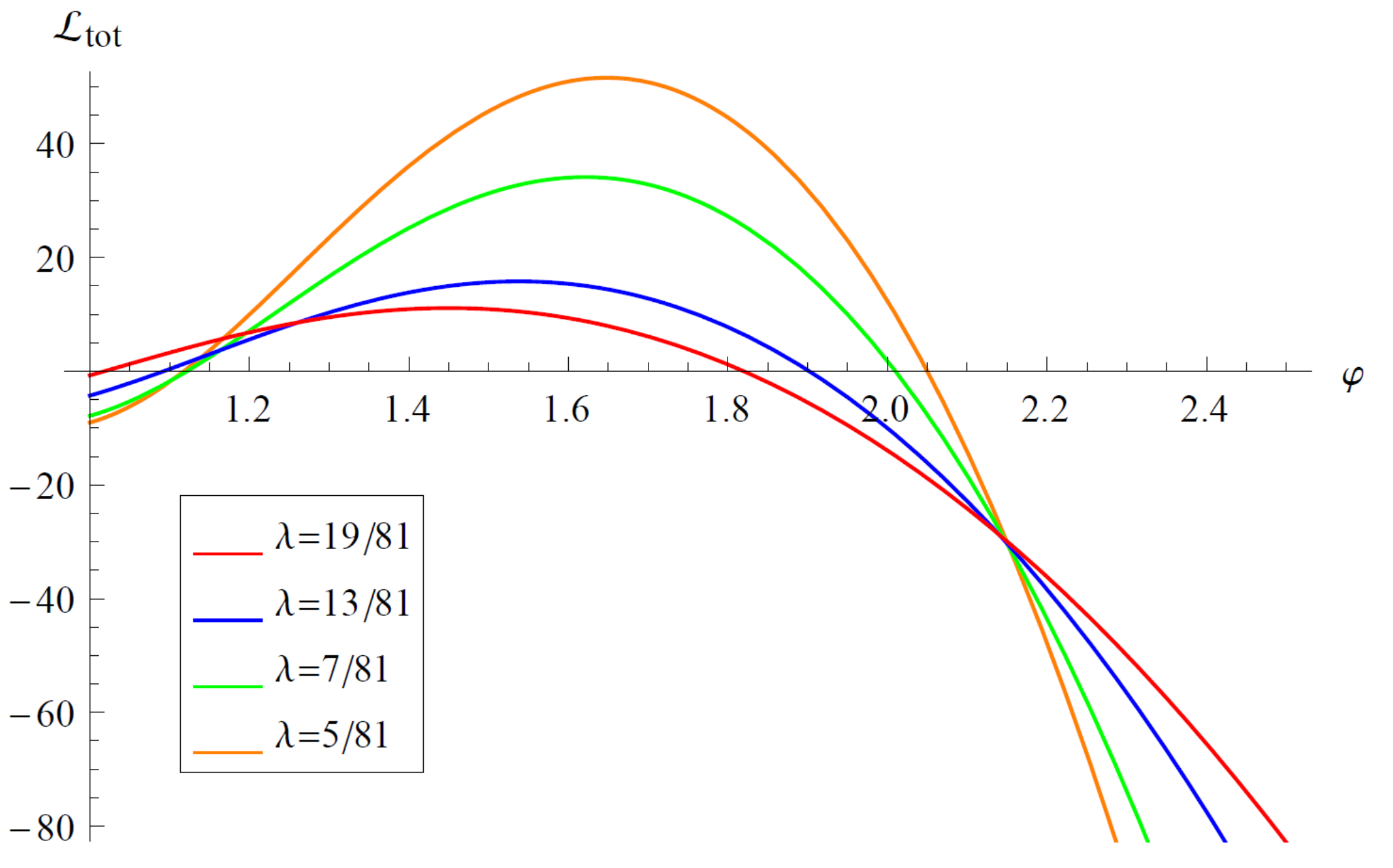}
\caption{{\it The plot for $\mathcal{L}_{tot}$ as a function of $\varphi$ for various $\lambda$. For any given $0<\lambda\leq 19/81$, the Lagrangian density always has two real roots in the black hole interior.}}
\label{LL3}\end{figure}
Evaluating the Lagrangian density for the solution yields
\bea \mathcal{L}_{tot}&=&
-\fft{4}{3\ell^2\lambda\varphi^5}\Big[15\varphi^7-40\varphi^6+3(45\lambda-7)\varphi^5+110(1-3\lambda)\varphi^4   \nn\\
&&\qquad\qquad\,\,\,-60(1-3\lambda)\varphi^3-20(1-3\lambda)^2\varphi+16(1-3\lambda)^2 \Big]\,.
\eea
It turns out that for any given $\lambda$ in the allowed region, the Lagrangian density always has two real roots in the black hole interior, see Fig.\ref{LL3}. The larger root corresponds to a local maxima for the growth rate of complexity. Of course, this case is not covered by our conditions (\ref{condition1}-\ref{condition2}). We turn to study the time evolution of complexity by adopting numerical approach. The result is presented in Fig.\ref{LL3cp}. It is easy to see that the complexity growth rate exceeds the Lloyd bound around the local maxima for any given $0<\lambda\leq 19/81$. For a smaller $\lambda$, the height of the local maximal growth rate becomes smaller. In particular, for a sufficiently small $\lambda$, we find
\be \fft{d\mathcal{C}_{max}}{dt}=\fft{2M}{\pi\hbar}\Big(1.0035+0.492 \lambda+\mathcal{O}(\lambda^2) \Big)>\fft{2M}{\pi\hbar} \,.\ee
Thus, we safely conclude that for any given $\lambda$ in the allowed region, our conjectured bound is always violated.

\section{An alternative proposal}\label{altenateproposal}

It was shown in \cite{Lehner:2016vdi,Carmi:2017jqz} that for ``Complexity=Action" (CA) proposal, the boundary action at the corner where two null boundaries of WDW patch intersect significantly contributes to the time dependence of complexity. Surprisingly, the growth rate of the joint action at late times exactly equals to the product $T S$. This has been shown for certain black holes in Einstein's gravity \cite{Fan:2018wnv,Lehner:2016vdi}, Lovelock gravity \cite{Cano:2018aqi}, $F(R)$ gravity and critical gravity \cite{Jiang:2018sqj}, respectively. In the following, we shall first prove that it is valid to generally static black holes in higher derivative gravities.

\subsection{Joint action growth at late times}\label{A1}
The joint action was derived from action principle for gravitational theories defined on spacetimes with nonsmooth boundaries \cite{Lehner:2016vdi}. Recently, the result was generalized to higher derivative gravities in \cite{Jiang:2018sqj,Cano:2018ckq}. One has
\be S_{joint}\equiv \fft{1}{2\pi}\int_{C_{n-2}}d\Omega_{n-2} \,s \,a \,.\ee
Here $s$ is the Wald entropy density function
\be s=-2\pi\, \fft{\partial\mathcal{L}}{\partial R_{\mu\nu\rho\sigma}}\,\varepsilon_{\mu\nu}\varepsilon_{\rho\sigma} \,,\ee
and $a$ is the standard corner term for Einstein's gravity. For null-null joint, one has \cite{Lehner:2016vdi}
\be a\equiv \epsilon\log{|\ft 12 k_1\cdot k_2|}\,,\qquad \epsilon=-\mathrm{sign}(k_1\cdot k_2)\,\mathrm{sign}(\hat{k}_1\cdot k_2) \,,\ee
\begin{figure}[htbp]
  \centering
  	\subfigure{\includegraphics[width=3in]{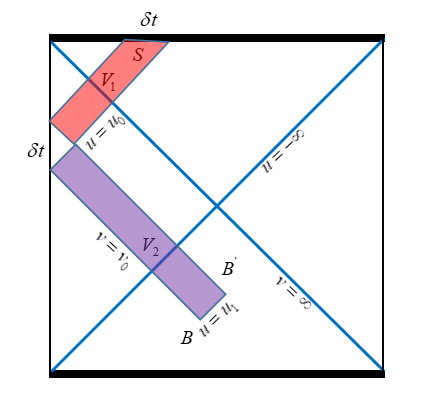}}
  	\caption{The action growth of the WDW patch of an uncharged black hole. In the time process $t_L\rightarrow t_L+\delta t$, the patch loses a sliver $V_2$ (Purple) and gains another one $V_1$ (Red). }
\label{wdw} \end{figure}
where $k_i$ are outward directed normal vectors of the dual null boundaries and $\hat{k}_i$ are auxiliary null vectors defined in the tangent space of the boundaries, orthogonal to the joint and pointing outward from the boundary regions. For static black holes, the corners are determined by codimension-2 two surfaces $t=\mathrm{const}\,,r=\mathrm{const}$. The joint action simplifies to
\be S_{joint}=\fft{1}{2\pi}\,S(r)a(r) \,.\ee

 For an infinitesimal time process $t_L\rightarrow t_L+\delta t$ shown in Fig.\ref{wdw}, the growth of the joint action takes the form of
\be\label{rm}
\delta S_{joint}=\frac{1}{2\pi }\oint_{B^{\prime}} d\Omega_{n-2}\,s\,a-\frac{1}{2\pi}\oint_{B} d\Omega_{n-2}\,s\,a\,.
\ee
The relevant null normals can be written as
\bea
k_\mu&=&-c\partial_\mu v=-c\partial_\mu(t-r^{\ast})\nn\\
\bar{k}_\mu&=&\bar{c}\partial_\mu u=\bar c\partial_\mu (t+r^{\ast})
\eea
where $c$ and $\bar{c}$ are arbitary positive constants which can be fixed by implementing the asymptotic normalizations $k\cdot \hat{t}_L=-c$ and $\bar{k}\cdot\hat{t}_R=-\bar{c}$, where $\hat{t}_{L,R}$ are the asymptotic Killing vectors on the left and right boundaries, respectively. Without loss of generality, we further choose $c=\bar{c}$. With these choices, we have $k\cdot\bar{k}=-2c^2/h$, so that
\bea
a=-\log\Big(\frac{h}{c^2}\Big)\,.
\eea
We deduce
\bea\label{growthjoint}
\delta S_{joint}=\frac{1}{2\pi }\Big( S(r_{B'})a(r_{B^{\prime}})-S(r_B)a(r_{B}) \Big)\,.
\eea
However, since the variation between $r_{B}$ and $r_{B^{\prime}}$ is very small, we can perform a Taylor expansion for $a(r)$ around $r=r_B$. Note that the displacement is in the direction of the $v$-axis. We have $du=0$, $dv=\delta t$, and $dr=-\frac{1}{2}\sqrt{hf}\delta t$. This gives rise to
\bea
&&S(r_{B'})-S(r_B)=-\ft{\delta t}{2}\, S'(r_B)h(r_B)/w(r_B)\,,\nn\\
&&a(r_{B^{\prime}})-a(r_{B})=\ft{\delta t}{2}\,h^{\prime}(r_B)/w(r_B)\,.
\eea
Substituting this into (\ref{growthjoint}), one finds at late times ($r_B$ approaches $r_h$, $f$ and $h$ approach zero)
\be\label{jt}
\fft{dS_{joint}}{dt}=T S\,.
\ee
Note that the result does not depend on the asymptotic normalization constants.

\subsection{An alternative proposal and its time dependence}
Based on the above result, one may propose
\be \mathcal{C}= \fft{\alpha}{\pi \hbar}\,S_{joint}  \,,\ee
where $\alpha$ is the same coefficient introduced in (\ref{proposal1}). Clearly, at late times this alternative proposal gives exactly the same result as the first proposal (\ref{proposal1}). However, it turns out that in this case, our conjectured upper bound is always violated in the full time evolution. To show this, we recall that in Fig.\ref{tfd} the joint evolving with time is the one at the position $r=r_m$. Hence,
\bea
S_{joint}&=&\mathrm{const}+\fft{1}{2\pi}S(r_m)a(r_m) \nn\\
&=&\mathrm{const}-\fft{1}{2\pi}S(r_m)\log{\Big(\fft{|h(r_m)|}{c^2} \Big)}\,.
\eea
It is straightforward to show that
\bea \fft{dS_{joint}}{dt}&=&\fft{h'(r_m)S(r_m)}{4\pi w(r_m)}-\fft{S'(r_m)}{4\pi w(r_m)} |h(r_m)|\log{\Big(\fft{|h(r_m)|}{c^2} \Big)}\,,\eea
On the other hand, at late times the position $r_m$ behaves as
\be r_m=r_h-c_m\, e^{-2\pi T t}+\cdots \,,\ee
where $c_m$ is a positive constant. It follows that around late times the complexity growth rate behaves as
\be \fft{d\mathcal{C}}{dt}=\fft{\alpha }{\pi\hbar}\Big(TS+2\pi c_m\, T S'(r_h)\,T t\, e^{-2\pi T t}+\cdots \Big) \,.\ee
 It is easily seen that the late time rate of change of complexity is always approached from above. Thus, unlike our first proposal, the complexity always violates the conjectured upper bound\footnote{In fact, the late time behavior of the joint action is also a major reason that why CA proposal does not respect the Lloyd bound. }.

\section{Conclusions}
In this paper, we propose that the complexity for a holographic state is dual to a simple gravitational object defined in the bulk of WDW patch (see Eq.(\ref{proposal1})), of which the growth rate at late times is equal to temperature times black hole entropy. Remarkably, thermodynamics of AdS black holes guarantees that for this new proposal the complexity growth rate at late times always saturates the Lloyd bound (and its proper generalisations for charged systems) at high temperature limit. This is universal to generally static AdS black holes. In particular, for AdS planar black holes, the result holds at lower temperatures as well owing to an extra scaling symmetry of the solutions.

We conjecture that the complexity growth rate should be bounded above as $d\mathcal{C}/dt\leq \alpha T S/\pi\hbar$ or $d\mathcal{C}/dt\leq \alpha \big(T_+ S_+-T_-S_-\big)/\pi\hbar$ for black holes with an inner horizon, where $\alpha$ is an overall coefficient (to connect the bound to the Lloyd bound at high temperatures, we choose it to be $\alpha=\fft{2(D-2)}{D-1}$, where $D$ denotes spacetime dimension). To test the conjecture, we study the time dependence of complexity for holographic states without perturbations. We show that the bound holds for holographic theories dual to Einstein's gravity coupled to matter fields which obey strong and weak energy conditions. However, with stringy corrections, the bound may be violated. This is ensured for black holes in third order Lovelock gravity.

In addition, we are aware of that the growth rate of the joint action of WDW patch gives the same result at late times as our new proposal. However, a major difference is it always violates the conjectured upper bound around late times.

\section*{Acknowledgments}
We are grateful to Hugo Marrochio for useful comments and discussions. Z.Y. Fan is supported in part by the National Natural Science Foundations of China (NNSFC) with Grant No. 11805041, No. 11873025 and No. 11575270. M. Guo is supported in part by NNSFC Grants No.11775022 and No.11375026 and also supported by the China Scholarship Council. M. Guo also thanks the Perimeter Institute \lq\lq{}Visiting Graduate Fellows\rq\rq{} program. Research at Perimeter Institute is supported in part by the Government of Canada through the Department of Innovation, Science and Economic Development and by the Province of Ontario through the Ministry of Research, Innovation and Science.\\

\appendix

\section{Smarr formula at high temperature limit}\label{smarrhightem}
According to scaling dimensional argument, the Smarr formula for charged AdS black holes is given by
\be\label{appsmarr} M-\mu Q=\ft{D-2}{D-3}\,T S-\ft{2}{D-3}\,P V_{ther} \,.\ee
Here for simplicity, we have assumed that all the higher order gravitational coupling constants as well as those characterizing self-interactions of matter fields in the Lagrangian are related to the cosmological constant. By generalizing the Wald-Iyer formalism, it was established in \cite{Fan:2018wnv} that the product $P V_{ther}$ is equal to the rate of change of non-derivative gravitational action defined on WDW patch. One has
\be\label{PV1} P V_{ther}=-\fft{\omega_{D-2}}{16\pi G}\int_{\epsilon}^{r_h}\mathrm{d}r\,\sqrt{-\bar g}\,V(\phi) \,,\ee
where $V(\phi)$ is the scalar potential which has a small $\phi$ expansion as $V=2\Lambda+\ft 12 m^2\phi^2+\cdots$. In high temperature limit $r_h\rightarrow \infty$, the dominant contribution to the integral comes from the region close to the event horizon. One has $V=2\Lambda+\cdots$ and
\be \sqrt{-\bar g}=r^{D-2}\sqrt{\ft{h(r)}{f(r)}}=r^{D-2}\Big(1+O(r-r_h) \Big) \,,\ee
where in the second ``=", we have set $h'(r_h)=f'(r_h)$ by scaling the time coordinate. Hence, to leading order one finds
\be P V_{ther}\simeq \fft{(D-2)\omega_{D-2}r_h^{D-1}}{16\pi G\ell^2} \,.\ee
On the other hand, one has
\be T S=\fft{\omega_{D-2}r_h^{D-2}}{16\pi G}\,f'(r_h) \,.\ee
Though the metric function $f(r)$ will be different for different black holes, we only need its large-$r$ expansion
\be f=r^2\ell^{-2}+k-\fft{16\pi}{(D-2)\omega_{D-2}}\fft{G M}{r^{D-3}}+\sum_{\sigma_i>0} \fft{c_i}{r^{\sigma_i}}\,.\ee
One finds
\be f'(r_h)=\fft{(D-1)r_h}{\ell^2}+\fft{(D-3)k}{r_h}+\sum_{\sigma_i>0}\fft{(D-3-\sigma_i)c_i}{r_h^{\sigma_i+1}} \,.\ee
It follows that at leading order
\be TS\simeq \fft{(D-1)\omega_{D-1}r_h^{D-1}}{16\pi G\ell^2} \,.\ee
Therefore in high temperature limit
\be P V_{ther}\simeq \ft{D-2}{D-1}\,T S \,.\ee
By plugging it into the Smarr formula (\ref{appsmarr}), one immediately finds
\be M-\mu Q\simeq \ft{D-2}{D-1}\,T S \,.\ee
Moreover, for charged black holes with inner horizons, the relation (\ref{PV1}) is replaced by \cite{Fan:2018wnv}
\be\label{PV2} P \big(V_+-V_-\big)=-\fft{\omega_{D-2}}{16\pi G}\int_{r_-}^{r_+}\mathrm{d}r\,\sqrt{-\bar g}\,V(\phi) \,.\ee
It follows that to leading order
\be P(V_+-V_-)\simeq \ft{(D-2)\omega_{D-2}}{16\pi G\ell^2}\,\big(r_+^{D-1}-r_-^{D-1} \big)
\simeq \ft{D-2}{D-1}\,\big(T_+S_+-T_-S_- \big) \,,\ee
so that
\be \big(M-\mu Q \big)_+-\big(M-\mu Q \big)_-\simeq  \ft{D-2}{D-1}\,\big(T_+S_+-T_-S_- \big)\,.\ee

\section{Action growth for general higher derivative gravities}\label{CAHDG}
In ``Complexity=Action" proposal \cite{Brown:2015bva,Brown:2015lvg}, the action growth of WDW patch plays a central role in studying properties of holographic complexity. Despite that it has been extensively studied in literatures, it still deserves further investigations for action growth rate at late times for generally static black holes in higher derivative gravities. We consider gravitational theories of the type
\be S_{bulk}=\int_{\mathcal{M}} d^Dx \sqrt{-g}\,\mathcal{L}(g_{\mu\nu}; R_{\mu\nu\rho\sigma} )  \,.\ee
The generalized Gibbons-Hawking boundary term is given by \cite{Deruelle:2009zk}
\be  S_{surf}=2\int_{\partial \mathcal{M}}d\Sigma_{D-1}\,2\fft{\partial\mathcal{L}}{\partial R_{\mu\sigma\nu\rho}}\,n_\sigma n_\rho K_{\mu\nu} \,,\ee
where $n^\mu$ is the normal vector of the time-like hypersurface $\partial \mathcal{M}$ and $K_{\mu\nu}$ is its second fundamental form.

In an infinitesimal time process $t_L\rightarrow t_L+\delta t$ shown in Fig.\ref{wdw}, the action growth $\delta S=S(t_L+\delta t)-S(t_L)$ takes the form of
\be\label{rmaction}
\delta S=S_{V_1}-S_{V_2}-S_{surf}(r=\epsilon)+S_{joint}(r_{B'})-S_{joint}(r_B)\,,
\ee
where $V_1$ and $V_2$ denote the upper and lower slivers in Fig.\ref{wdw}, respectively. First, we shall prove at late times
\be\label{bulkindentity} S_{V_1}-S_{V_2}=\dot{S}_{bulk}\,\delta t \,,\ee
where $\dot{S}_{bulk}$ is the variation of bulk gravitational action with boundary time. For this purpose, we introduce the null coordinates $u$ and $v$ defined by
\bea
u&=&t+r^*\,,\nn\\
v&=&t-r^*\,.
\eea
The metric can be written as
\bea
ds^2=-hdu^2+2w dudr+r^2d\Omega_{D-2\,,k}^2\,,
\eea
in the ingoing coordinate and
\bea
ds^2=-hdv^2-2w dvdr+r^2d\Omega_{D-2\,,k}^2\,,
\eea
in the outgoing coordinate, respectively. In addition, note that $dt\wedge dr=du\wedge dr=dv\wedge dr$, we have a simple and useful relation
\be
\int d^Dx\,\sqrt{-g}=\int d\Omega_{D-2}\,d\tau dr\,\sqrt{-g}=\omega_{D-2}\int d\tau dr\,\sqrt{-\bar g}\,,
\ee
where $\tau$ collectively denotes the various time coordinates $t,u,v$.

For the first rectangle $V_1$, one has
\bea
S_{V_1}&=&\int_{u_0}^{u_0+\delta t}du\int d\Omega_{D-2} \int^{\lambda(u)}_{\epsilon}dr\sqrt{-g}\mathcal{L}\,,\nn\\
&=&\omega_{D-2}\int_{u_0}^{u_0+\delta t}du\, \int^{\lambda(u)}_{\epsilon}dr\sqrt{-g}\mathcal{L}\,,
\eea
where in the first \lq\lq{}=\rq\rq{}, we denote $r=\lambda(u)$ to describe the null hypersurface $v=v_0+\delta t$. The function $\lambda(u)$ can be solved by the equation $r^*\big(\lambda(u)\big)=\frac{1}{2}(u-v_0-\delta t)$. Similarly, for the rectangle $V_2$, we have
\bea
S_{V_2}&=&\int_{v_0}^{v_0+\delta t}dv\int d\Omega_{D-2}\int^{\lambda_0(v)}_{\lambda_1(v)}dr\sqrt{-g}\mathcal{L}\,,\nn\\
&=&\omega_{D-2}\int_{v_0}^{v_0+\delta t}dv\, \int^{\lambda_0(v)}_{\lambda_1(v)}dr\sqrt{-g}\mathcal{L}\,.
\eea
Here we define $r=\lambda_{0,1}(v)$ to describe the null hypersurfaces $u=u_{0\,,1}$. They are determined by $r^*\big(\lambda_{0\,,1}(v) \big)=\frac{1}{2}(u_{0\,,1}-v)$.
Then changing the variables $u=u_0+v_0+\delta t-v$ in the integral of $S_{V_1}$, we find
\bea
S_{V_1}&=&\omega_{D-2}\int_{v_0}^{v_0+\delta t}dv \int^{\lambda_0(v)}_{\epsilon}dr\sqrt{-g}\mathcal{L}\,,
\eea
where we have identified $\lambda_0(v)=\lambda(u)$ since they both describe a same radii at which the null boundaries $u=u_0$ and $v=v_0+\delta t$ intersect.
Combining the above results, we deduce
\bea
S_{V_1}-S_{V_2}&=&\omega_{D-2}\int_{v_0}^{v_0+\delta t}dv \,\int^{\lambda_1(v)}_{\epsilon}dr\sqrt{-g}\mathcal{L}\,.
\eea
Considering the radii $r=\lambda_1(v)$ varies from $r_B$ to $r_{B^{\prime}}$ as $v$ increases from $v_0$ to $v_0+\delta t$, we find $r_{B^{\prime}}=r_{B}+O(\delta t)$ since the variation of the radius is very small. In particular, at late times, $r_{B}$ approaches $r_h$. So we arrive at
\be\label{bt}
S_{V_1}-S_{V_2}=\omega_{D-2}\int_{v_0}^{v_0+\delta t}dv \,\int^{r_h}_{\epsilon}dr\sqrt{-g}\mathcal{L}=\dot{S}_{bulk}\,\delta t\,.
\ee
It should be emphasized that this equality is valid to any gravitational object defined in the bulk of WDW patch.

On the other hand, according to Eq.(\ref{jt}), at late times the growth rate of the joint action equals to $T S$. However, in \cite{Huang:2016fks}, it was also shown that
\be \fft{dS_{surf}}{dt}\Big|_{r\rightarrow r_h^+}=T S \,.\ee
Thus, at late times
\be \fft{dS_{joint}}{dt}=\fft{dS_{surf}}{dt}\Big|_{r\rightarrow r_h^+}\,.\ee
Combing all the above results, we obtain
\be \fft{dS}{dt}=\fft{dS_{bulk}}{dt}+\fft{dS_{surf}}{dt}\Big|_{r=r_h}-\fft{dS_{surf}}{dt}\Big|_{r=\epsilon} \,.\ee
This is valid to generally static black holes in higher derivative gravities. The result covers previous results published in literatures \cite{Fan:2018wnv,Cai:2016xho,Lehner:2016vdi,Huang:2016fks,Cano:2018aqi,Jiang:2018sqj}.

\section{Deriving the conditions (\ref{condition31}\,,\ref{condition32})}\label{appcondition}

\begin{figure}[htbp]
  \centering
  	\subfigure{\includegraphics[width=2.5in]{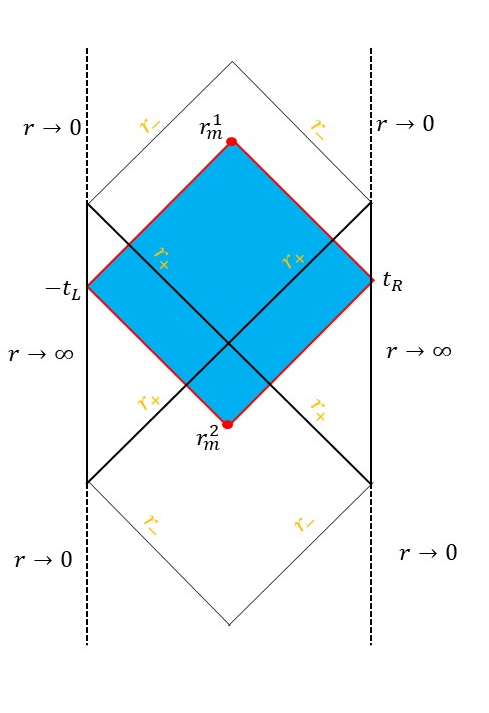}}
  	\caption{The WDW patch of an AdS black hole with an inner horizon. There are always two joints $r_m^1\,,r_m^2$, evolving with time. }
\label{tfdc}
\end{figure}

For black holes with an inner horizon, the WDW patch (see Fig.\ref{tfdc}) has two joints, denoted by $r_m^1\,,r_m^2$ respectively. 
The evolution of the joints is determined by
\bea
t\equiv t_R+t_L=2r^*(r_m^1)=-2r^*(r_m^2)\,,
\eea
where $r_-\leq r_m^1, r_m^2\leq r_+$. Evaluating their time derivatives yields
\bea
\frac{dr_m^1}{dt}&=&\frac{1}{2}w(r_m^1)f(r_m^1)\leq0 \,,\nn\\
\frac{dr_m^2}{dt}&=&-\frac{1}{2}w(r_m^2)f(r_m^2)\geq0 \,.
\eea
On the other hand, the complexity growth rate is given by
\bea
\frac{d\mathcal{C}}{dt}&=&-\frac{\omega_{D-2}}{\pi\hbar}\int^{r_+}_{r_m^1}dr\sqrt{-\bar{g}}\,(\mathcal{L}_{grav}-2\rho)\nn\\
&&+\frac{\omega_{D-2}}{\pi\hbar}\int^{r_+}_{r_m^2}dr\sqrt{-\bar{g}}\,(\mathcal{L}_{grav}-2\rho)\nn\\
&=&-\frac{\omega_{D-2}}{\pi\hbar}\int^{r_m^2}_{r_m^1}dr\sqrt{-\bar{g}}\,(\mathcal{L}_{grav}-2\rho)\,.
\eea
Thus, taking one more derivative with respect to time, one finds
\be
\frac{d^2\mathcal{C}}{dt^2}=-\frac{\omega_{D-2}}{\pi\hbar}\sum_{i=1,2}\Big[\sqrt{-\bar{g}}(\mathcal{L}_{grav}-2\rho)\Big]_{r=r_m^i}|\dot{r}_m^i|\,.
\ee
It follows that under the condition (\ref{condition31}) the rate of change of complexity is a monotone increasing function of time and hence obeys the bound $d\mathcal{C}/dt\leq \alpha\big(T_+S_+-T_-S_- \big)/\pi\hbar$.

On the other hand, at late times, one has
\bea
r_m^1&=&r_-+c_-e^{2\pi T_- t}\,,\nn\\
r_m^2&=&r_+-c_+e^{-2\pi T_+ t}\,,
\eea
where $c_{\pm}$ are positive constants. One finds
\bea
\frac{d\mathcal{C}}{dt}&=&\fft{d\mathcal{C}}{dt}\Big|_{late}+\frac{\omega_{D-2}}{\pi\hbar}\left[\sqrt{-\bar{g}}(\mathcal{L}_{grav}-2\rho)\right]_{r=r_-}c_-e^{2\pi T_- t}\nn\\
&&+\frac{\omega_{D-2}}{\pi\hbar}\left[\sqrt{-\bar{g}}(\mathcal{L}_{grav}-2\rho)\right]_{r=r_+}c_+e^{-2\pi T_+ t}\,.
\eea
Therefore, as discussed above, under the condition (\ref{condition32}), the upper bound on complexity is again obeyed.


\begin{thebibliography}{100}
\bibitem{Margolus:1997ih}
  N.~Margolus and L.~B.~Levitin,
  {\it The Maximum speed of dynamical evolution,}
  Physica D {\bf 120}, 188 (1998) [quant-ph/9710043].

\bibitem{Lloyd}
S.~Lloyd, {\it Ultimate physical limits to computation},
Nature {\bf 406} (2000), no. 6799 10471054.


\bibitem{Jordan:2017vqh}
  S.~P.~Jordan,
  {\it Fast quantum computation at arbitrarily low energy,}
  Phys.\ Rev.\ A {\bf 95}, no. 3, 032305 (2017)
  [arXiv:1701.01175 [quant-ph]].

\bibitem{Stanford:2014jda}
  D.~Stanford and L.~Susskind,
  {\it Complexity and Shock Wave Geometries,}
  Phys.\ Rev.\ D {\bf 90}, no. 12, 126007 (2014)
  [arXiv:1406.2678 [hep-th]].

\bibitem{Brown:2015bva}
  A.~R.~Brown, D.~A.~Roberts, L.~Susskind, B.~Swingle and Y.~Zhao,
  {\it Holographic Complexity Equals Bulk Action?,}
  Phys.\ Rev.\ Lett.\  {\bf 116}, no. 19, 191301 (2016)
  [arXiv:1509.07876 [hep-th]].

\bibitem{Brown:2015lvg}
  A.~R.~Brown, D.~A.~Roberts, L.~Susskind, B.~Swingle and Y.~Zhao,
  {\it Complexity, action, and black holes,}
  Phys.\ Rev.\ D {\bf 93}, no. 8, 086006 (2016)
  [arXiv:1512.04993 [hep-th]].


\bibitem{Deruelle:2009zk}
  N.~Deruelle, M.~Sasaki, Y.~Sendouda and D.~Yamauchi,
  {\it Hamiltonian formulation of f(Riemann) theories of gravity,}
  Prog.\ Theor.\ Phys.\  {\bf 123}, 169 (2010)
  [arXiv:0908.0679 [hep-th]].


\bibitem{Couch:2016exn}
  J.~Couch, W.~Fischler and P.~H.~Nguyen,
  {\it Noether charge, black hole volume, and complexity,}
  JHEP {\bf 1703}, 119 (2017)
  [arXiv:1610.02038 [hep-th]].

\bibitem{Fan:2018wnv}
  Z.~Y.~Fan and M.~Guo,
  {\it On the Noether charge and the gravity duals of quantum complexity,}
  JHEP {\bf 1808}, 031 (2018)
  [arXiv:1805.03796 [hep-th]].


\bibitem{Fan:2018xwf}
  Z.~Y.~Fan and M.~Guo,
  {\it Holographic complexity under a global quantum quench,}
  arXiv:1811.01473 [hep-th].


\bibitem{Moosa:2017yiz}
  M.~Moosa,
  {\it Divergences in the rate of complexification,}
  Phys.\ Rev.\ D {\bf 97}, no. 10, 106016 (2018)
  [arXiv:1712.07137 [hep-th]].


\bibitem{HosseiniMansoori:2017tsm}
  S.~A.~Hosseini Mansoori and M.~M.~Qaemmaqami,
  {\it Complexity Growth, Butterfly Velocity and Black hole Thermodynamics,}
  arXiv:1711.09749 [hep-th].


\bibitem{Mahapatra:2018gig}
  S.~Mahapatra and P.~Roy,
  {\it On the time dependence of holographic complexity in a dynamical Einstein-dilaton model,}
  JHEP {\bf 1811}, 138 (2018)
  [arXiv:1808.09917 [hep-th]].




\bibitem{Chapman:2016hwi}
  S.~Chapman, H.~Marrochio and R.~C.~Myers,
  {\it Complexity of Formation in Holography,}
  JHEP {\bf 1701}, 062 (2017)
  [arXiv:1610.08063 [hep-th]].


\bibitem{Carmi:2016wjl}
  D.~Carmi, R.~C.~Myers and P.~Rath,
  {\it Comments on Holographic Complexity,}
  JHEP {\bf 1703}, 118 (2017)
  [arXiv:1612.00433 [hep-th]].



\bibitem{Kim:2017lrw}
  R.~Q.~Yang, C.~Niu and K.~Y.~Kim,
  {\it Surface Counterterms and Regularized Holographic Complexity,}
  JHEP {\bf 1709}, 042 (2017)
  [arXiv:1701.03706 [hep-th]].



\bibitem{Yang:2016awy}
  R.~Q.~Yang,
  {\it Strong energy condition and complexity growth bound in holography,}
  Phys.\ Rev.\ D {\bf 95}, no. 8, 086017 (2017)
  [arXiv:1610.05090 [gr-qc]].



\bibitem{Moosa:2017yvt}
  M.~Moosa,
  {\it Evolution of Complexity Following a Global Quench,}
  JHEP {\bf 1803}, 031 (2018)
  [arXiv:1711.02668 [hep-th]].


\bibitem{Swingle:2017zcd}
  B.~Swingle and Y.~Wang,
  {\it Holographic Complexity of Einstein-Maxwell-Dilaton Gravity,}
  arXiv:1712.09826 [hep-th].


\bibitem{Alishahiha:2018tep}
  M.~Alishahiha, A.~Faraji Astaneh, M.~R.~Mohammadi Mozaffar and A.~Mollabashi,
  {\it Complexity Growth with Lifshitz Scaling and Hyperscaling Violation,}
  JHEP {\bf 1807}, 042 (2018)
  [arXiv:1802.06740 [hep-th]].


\bibitem{An:2018xhv}
  Y.~S.~An and R.~H.~Peng,
  {\it Effect of the dilaton on holographic complexity growth,}
  Phys.\ Rev.\ D {\bf 97}, no. 6, 066022 (2018)
  [arXiv:1801.03638 [hep-th]].


\bibitem{Jiang:2018pfk}
  J.~Jiang,
  {\it Action growth rate for a higher curvature gravitational theory,}
  Phys.\ Rev.\ D {\bf 98}, no. 8, 086018 (2018)
  [arXiv:1810.00758 [hep-th]].


\bibitem{Kim:2017qrq}
  R.~Q.~Yang, C.~Niu, C.~Y.~Zhang and K.~Y.~Kim,
  {\it Comparison of holographic and field theoretic complexities for time dependent thermofield double states,}
  JHEP {\bf 1802}, 082 (2018)
  [arXiv:1710.00600 [hep-th]].


\bibitem{Yang:2019gce}
  R.~Yang, H.~S.~Jeong, C.~Niu and K.~Y.~Kim,
  {\it Complexity of Holographic Superconductors,}
  arXiv:1902.07586 [hep-th].



\bibitem{Guo:2019vni}
  H.~Guo, X.~M.~Kuang and B.~Wang,
  {\it Note on holographic entanglement entropy and complexity in St$\ddot{u}$ckelberg superconductor,}
  arXiv:1902.07945 [hep-th].





















\bibitem{Susskind:2014rva}
  L.~Susskind,
  {\it Computational Complexity and Black Hole Horizons,}
  Fortsch.\ Phys.\  {\bf 64}, 24 (2016)
  [arXiv:1402.5674 [hep-th]].

\bibitem{wald1}
  R.M.~Wald,
{\it Black hole entropy is the Noether charge},
Phys.\ Rev.\ D {\bf 48}, 3427 (1993), gr-qc/9307038.

\bibitem{wald2}  V.~Iyer and R.M.~Wald,
{\it Some properties of Noether charge and a proposal for
dynamical black hole entropy,}
Phys.\ Rev.\ D {\bf 50}, 846 (1994),  gr-qc/9403028.


\bibitem{Fan:2018qnt}
  Z.~Y.~Fan,
  {\it Note on the Noether charge and holographic transports,}
  Phys.\ Rev.\ D {\bf 97}, no. 6, 066013 (2018)
  [arXiv:1801.07870 [hep-th]].


\bibitem{Liu:2015tqa}
  H.~S.~Liu, H.~Lu and C.~N.~Pope,
  {\it Generalized Smarr formula and the viscosity bound for Einstein-Maxwell-dilaton black holes,}
  Phys.\ Rev.\ D {\bf 92}, 064014 (2015)
  [arXiv:1507.02294 [hep-th]].




\bibitem{Lu:2014maa}
  H.~Lu, C.~N.~Pope and Q.~Wen,
  {\it Thermodynamics of AdS Black Holes in Einstein-Scalar Gravity,}
  JHEP {\bf 1503}, 165 (2015)
  [arXiv:1408.1514 [hep-th]].



\bibitem{Cai:2016xho}
  R.~G.~Cai, S.~M.~Ruan, S.~J.~Wang, R.~Q.~Yang and R.~H.~Peng,
  {\it Action growth for AdS black holes,}
  JHEP {\bf 1609}, 161 (2016)
  [arXiv:1606.08307 [gr-qc]].


\bibitem{Lehner:2016vdi}
  L.~Lehner, R.~C.~Myers, E.~Poisson and R.~D.~Sorkin,
  {\it Gravitational action with null boundaries,}
  Phys.\ Rev.\ D {\bf 94}, no. 8, 084046 (2016)
  [arXiv:1609.00207 [hep-th]].

\bibitem{Huang:2016fks}
  H.~Huang, X.~H.~Feng and H.~Lu,
  {\it Holographic Complexity and Two Identities of Action Growth,}
  Phys.\ Lett.\ B {\bf 769}, 357 (2017)
  [arXiv:1611.02321 [hep-th]].


\bibitem{Cano:2018aqi}
  P.~A.~Cano, R.~A.~Hennigar and H.~Marrochio,
  {\it Complexity Growth Rate in Lovelock Gravity,}
  Phys.\ Rev.\ Lett.\  {\bf 121}, no. 12, 121602 (2018)
  [arXiv:1803.02795 [hep-th]].

\bibitem{Jiang:2018sqj}
  J.~Jiang and H.~Zhang,
  {\it Surface term, corner term, and action growth in F(Riemann) gravity theory,}
  arXiv:1806.10312 [hep-th].



\bibitem{Jiang:2019fpz}
  J.~Jiang and X.~W.~Li,
  {\it Modified ``complexity equals action" conjecture,}
  arXiv:1903.05476 [hep-th].


\bibitem{Feng:2018sqm}
  X.~H.~Feng and H.~S.~Liu,
  {\it Holographic Complexity Growth Rate in Horndeski Theory,}
  Eur.\ Phys.\ J.\ C {\bf 79}, no. 1, 40 (2019)
  [arXiv:1811.03303 [hep-th]].


\bibitem{Alishahiha:2017hwg}
  M.~Alishahiha, A.~Faraji Astaneh, A.~Naseh and M.~H.~Vahidinia,
  {\it On complexity for F(R) and critical gravity,}
  JHEP {\bf 1705}, 009 (2017)
  [arXiv:1702.06796 [hep-th]].


\bibitem{Carmi:2017jqz}
  D.~Carmi, S.~Chapman, H.~Marrochio, R.~C.~Myers and S.~Sugishita,
  {\it On the Time Dependence of Holographic Complexity,}
  JHEP {\bf 1711}, 188 (2017)
  [arXiv:1709.10184 [hep-th]].


\bibitem{Brigante:2007nu}
  M.~Brigante, H.~Liu, R.~C.~Myers, S.~Shenker and S.~Yaida,
  {\it Viscosity Bound Violation in Higher Derivative Gravity,}
  Phys.\ Rev.\ D {\bf 77}, 126006 (2008)
  [arXiv:0712.0805 [hep-th]].

\bibitem{Camanho:2009vw}
  X.~O.~Camanho and J.~D.~Edelstein,
  {\it Causality constraints in AdS/CFT from conformal collider physics and Gauss-Bonnet gravity,}
  JHEP {\bf 1004}, 007 (2010)
  [arXiv:0911.3160 [hep-th]].


\bibitem{deBoer:2009gx}
  J.~de Boer, M.~Kulaxizi and A.~Parnachev,
  {\it Holographic Lovelock Gravities and Black Holes,}
  JHEP {\bf 1006}, 008 (2010)
  [arXiv:0912.1877 [hep-th]].


\bibitem{Camanho:2009hu}
  X.~O.~Camanho and J.~D.~Edelstein,
  {\it Causality in AdS/CFT and Lovelock theory,}
  JHEP {\bf 1006}, 099 (2010)
  [arXiv:0912.1944 [hep-th]].


\bibitem{Cano:2018ckq}
  P.~A.~Cano,
  {\it Lovelock action with nonsmooth boundaries,}
  Phys.\ Rev.\ D {\bf 97}, no. 10, 104048 (2018)
  [arXiv:1803.00172 [gr-qc]].






\end{thebibliography}
\end{document}